\documentclass[prd,superscriptaddress,nofootinbib,amsmath,amssymb,aps,11pt]{revtex4-2}

\usepackage{bm}
\usepackage{amsfonts}
\usepackage{latexsym}
\usepackage[latin1]{inputenc}
\usepackage{graphicx}
\usepackage{amsmath}
\usepackage{amssymb}
\usepackage{palatino}
\usepackage{mathpazo}
\usepackage{xcolor}
\usepackage{tikz}
\usetikzlibrary{shapes,arrows}

\usepackage{booktabs}
\usepackage{dcolumn}

\def\jnl@style{\it}
\def\aaref@jnl#1{{\jnl@style#1}}

\def\aaref@jnl#1{{\jnl@style#1}}

\def\aj{\aaref@jnl{AJ}}                   % Astronomical Journal
\def\apj{\aaref@jnl{ApJ}}                 % Astrophysical Journal
\def\apjl{\aaref@jnl{ApJ}}                % Astrophysical Journal, Letters
\def\apjs{\aaref@jnl{ApJS}}               % Astrophysical Journal, Supplement
\def\apss{\aaref@jnl{Ap\&SS}}             % Astrophysics and Space Science
\def\aap{\aaref@jnl{A\&A}}                % Astronomy and Astrophysics
\def\aapr{\aaref@jnl{A\&A~Rev.}}          % Astronomy and Astrophysics Reviews
\def\aaps{\aaref@jnl{A\&AS}}              % Astronomy and Astrophysics, Supplement
\def\mnras{\aaref@jnl{Mon.~Not.~Roy.~Astron.~Soc.}}             % Monthly Notices of the RAS
\def\prd{\aaref@jnl{Phys.~Rev.~D}}        % Physical Review D
\def\prc{\aaref@jnl{Phys.~Rev.~C}}        % Physical Review C
\def\prl{\aaref@jnl{Phys.~Rev.~Lett.}}    % Physical Review Letters
\def\qjras{\aaref@jnl{QJRAS}}             % Quarterly Journal of the RAS
\def\skytel{\aaref@jnl{S\&T}}             % Sky and Telescope
\def\ssr{\aaref@jnl{Space~Sci.~Rev.}}     % Space Science Reviews
\def\zap{\aaref@jnl{ZAp}}                 % Zeitschrift fuer Astrophysik
\def\nat{\aaref@jnl{Nature}}              % Nature
\def\aplett{\aaref@jnl{Astrophys.~Lett.}} % Astrophysics Letters
\def\apspr{\aaref@jnl{Astrophys.~Space~Phys.~Res.}} % Astrophysics Space Physics Research
\def\physrep{\aaref@jnl{Phys.~Rep.}}      % Physics Reports
\def\physscr{\aaref@jnl{Phys.~Scr}}       % Physica Scripta
\def\commat{\aaref@jnl{Comm.~Math.~Phys.}}              % Communications in Mathematical Physics
\def\science{\aaref@jnl{Science}}               % Science
\def\cqg{\aaref@jnl{Classical Quant.~Grav.}}            % Classical and Quantum Gravity
\def\jpcs{\aaref@jnl{JPCS}}                                     % Journal of Physics Conference Series
\def\ijmpd{\aaref@jnl{Int.~J.~Mod.~Phys.~D}}                    % International Journal of Modern Physics D
\def\grg{\aaref@jnl{Gen.~Relat.~Gravit.}}               % General Relativity and Gravitation
\def\rpp{\aaref@jnl{Rep.~Prog.~Phys.}}          % Reports on Progress in Physics
\def\npa{\aaref@jnl{Nucl.~Phys.~A}}        % Nuclear Physics A
\def\lrr{\aaref@jnl{Living Rev.~Rel.}}                   % Living reviews in relativity
\def\jcap{\aaref@jnl{J.~Cosmology Astropart.~Phys.}}    % Journal of cosmology and astroparticle physics
\def\rmp{\aaref@jnl{Rev.~Mod.~Phys.}}   %Reviews of modern physics

\allowdisplaybreaks[1]
\tikzstyle{block} = [rectangle, draw, text width=15em, text centered, rounded corners, minimum height=4em]
\tikzstyle{block_s} = [rectangle, draw, text width=10em, text centered, rounded corners, minimum height=4em]
\tikzstyle{line} = [draw, -latex']

\begin{document}
	
	\title{Tensor-Multiscalar Gravity: Equations of Motion to 2.5 post-Newtonian Order}
	
	\author{Oliver Sch\"on}
	\email{oliver.schoen@uni-tuebingen.de}
	\affiliation{Theoretical Astrophysics, Eberhard Karls University of T\"ubingen, T\"ubingen 72076, Germany}
	
	\author{Daniela D. Doneva}
	\email{daniela.doneva@uni-tuebingen.de}
	\affiliation{Theoretical Astrophysics, Eberhard Karls University of T\"ubingen, T\"ubingen 72076, Germany}
	
	%%%%%%%%%%%%%%%%%%%%%%%%%%%%%%%%%%%%  DATE  %%%%%%%%%%%%%%%%%%%%%%%%%%%%%%%%%%%%
	%\date{\today}
	
	\begin{abstract}
		In the present paper we take a step toward the generalization of the post-Newtonian formalism to tensor-multiscalar theories. These are theories where we have more than one scalar field being mediators of the gravitational interaction in addition to the spacetime metric. They are very natural extensions of Einstein's gravity allowing for the existence of new classes of compact objects and offering interesting phenomenology reaching far beyond the single scalar field theories. We calculate the expansion up to 2.5 post-Newtonian order in the near-zone using the so-called direct integration of the relaxed Einstein equations formalism and derive the equation of motion. This work is the first step toward the calculation of gravitational waveforms in tensor-multiscalar theories. 
	\end{abstract}
	
	\maketitle
	
	\section{INTRODUCTION}

		Testing General Relativity (GR) in the era of gravitational wave astrophysics has advanced a lot. There are still viable alternatives, though, mainly generalizations of GR  left to explore. One quite natural extension of GR is the introduction of a scalar field in addition to the metric tensor being an additional mediator of the gravitational interaction. We call those well-known generalizations of GR scalar-tensor theories (STT) \cite{fujii_maeda_2003,Damour:1992we}. Due to its simplicity the single scalar field case was much more widely considered in the literature. There is, however, no particular reason for appending GR by only one single scalar field. Even more, there are a number of motivations behind the idea that GR should be supplemented with multiple scalar fields related e.g. to higher dimensional gravity, string theories, etc. (see e.g. \cite{Damour:1994zq, Damour:1994ya, Albrecht:2001xt, Kainulainen:2004vk}). In addition, it is well known that different classes of alternative theories of gravity are mathematically equivalent to certain sectors of scalar-tensor theories that often offer the possibility of an easier and uniform treatment of these theories. Clearly, allowing for the existence of multiple scalar fields extends the possibilities for such analogies \cite{Rosa:2021lhc}. Taking additionally into account that those theories are mathematically well-defined and they are able to pass all known experimental and observational tests, makes the less explored class of tensor-multiscalar theories of gravitation (TMST) a viable and very interesting class of modified gravity. As a matter of fact, ones of the first seminal works on the topic of scalar-tensor theories by Damour and Esposito-Far\'ese considered the possibility for multiple scalar fields \cite{Damour:1992we, Damour:1995kt} and recently the $3+1$ formulation of the theory was developed in \cite{Horbatsch:2015bua}.
		
		What is very interesting about TMST from a theory point of view is that it is not just a mechanical addition of more scalar fields. Instead, these theories offer the possibility for the existence of completely new phenomena and solutions unseen in any other modified gravity theory until now.  The richness of the solution's spectrum is controlled by the choice of target space for the scalar fields equipped with a given metric, and the choice of the map $\varphi \colon spacetime \rightarrow target\; space$. Compact objects in TMST, including black holes, neutron stars and solitons, were studied in a number of papers \cite{Horbatsch:2015bua, Yazadjiev:2019oul, Doneva:2019krb, Doneva:2019ltb, Collodel:2019uns, Doneva:2020afj, Doneva:2020csi, Danchev:2020wnl, Collodel:2020gyp, Falcone:2020qqo, Kuan:2021yih, Sanchis-Gual:2021edp}. It was demonstrated that if one chooses the target space metric and the map to the target space in a nontrivial way, completely new types of solutions can exist, such as topological neutron stars \cite{Doneva:2019ltb}. Scalarization, that is a nonlinear scalar field development for sectors of the theory where the weak field regime coincides with GR \cite{Damour:1993hw}, was also considered in the context of TSMT \cite{Horbatsch:2015bua, Doneva:2020afj} and surprisingly it is possible to have scalarization with massless scalar fields leading to compact objects with zero scalar charge. This is contrary to all known scalarized solutions, both neutron stars and black holes \cite{Damour:1993hw, Doneva:2017bvd, Silva:2017uqg}, and suggests possible important deviations from the standard treatment of scalarization. Moreover, contrary to standard STT, black holes with scalar hair can exist in TMST \cite{Collodel:2020gyp}. All this calls for further development of TMST in the direction of studying possible astrophysical implications.

        Among the most promising test beds for alternative theories of gravity are binary mergers and especially their inspiral phase that is able to produce a strong signal-to-noise ratio. These phenomena are most prominently studied via a post-Newtonian (PN) approach \cite{Blanchet:2013haa, poisson2014gravity} that attracted considerable attention during the past decades due to its ability to produce fast and accurate enough waveforms. The recent results in GR include an analysis to 4 PN order \cite{Foffa:2012rn, Jaranowski:2012eb, Jaranowski:2013lca, Damour:2014jta, Bernard:2015njp, Bernard:2016wrg, Bernard:2017bvn, Marchand:2017pir} and 4.5 PN order \cite{Marchand:2016vox, Tucker:2021mvo}. The energy-flux was also studied in \cite{Blanchet:1995ez, Blanchet:1997jj, Blanchet:2001ax, Blanchet:2004ek} to 3.5 PN order. Gravitational waveforms were studied to 3.5 PN order \cite{Blanchet:2008je, Faye:2012we, Faye:2014fra}. Later the 5 PN order  \cite{Foffa:2019hrb, Blumlein:2019zku, Bini:2020wpo, Blumlein:2020pyo, Foffa:2020nqe} and the 6 PN order \cite{Bini:2020nsb,Bini:2020hmy} were considered as well. 
        
        One common school of thought is to calculate the PN expansion via a direct integration of the relaxed field equations (DIRE), pioneered by Epstein and Wagoner \cite{epstein1975post} and expanded by Wiseman, Will, Pati, Wang, and Mitchell \cite{Wiseman:1992dv, Will:1996zj, Will:1999dq, Pati:2000vt, Pati:2002ux, Will:2005sn, Wang:2007ntb, Mitchell:2007ea}. This series of papers establishes an equation of motion and radiation-reaction for binaries to 3.5 PN order in GR. Their method relies on weakening the standard field equations into a relaxed form as studied by Landau and Lifshitz \cite{Landau:1975pou}. This framework is understood as post-Minkowskian theory. The formalism allows one to rewrite the exact field equations as a set of ten flat, i.e. Minkowskian, wave equations together with imposing harmonic gauge conditions. Of course the source terms are not trivial as they are highly non-linear and convoluted. Via an iteration process \cite{Will:1999dq} one can expand the metric systematically and then integrate the wave equations using concepts such as retarded Green`s functions. This concept merges directly in the post-Newtonian formalism by incorporating both weak-field and slow-motion conditions $Gm/rc^2 \sim v^2/c^2 \ll 1$, where the characteristic mass, size, and velocity of the source are denoted by $m$, $r$, and $v$. Using geometric coordinate units, i.e. $G=c=1$, these conditions enable us to use the expansion parameter $\varepsilon \sim m/r \sim v^2$ to expand the metric fields $h^{\alpha \beta}:=\eta^{\alpha \beta}-\sqrt{-g}g^{\alpha \beta}$ for the Minkowski metric $\eta^{\alpha \beta}$ and spacetime metric $g^{\alpha \beta}$, $g=\det(g^{\alpha \beta})$. To integrate the wave equations for the $h^{\alpha \beta}$ fields, we decompose the past light-cone in a near-zone domain $\mathcal{N}$ and a wave-zone domain $\mathcal{W}$, such that $h^{\alpha \beta} = h_{\mathcal{N}}^{\alpha \beta} + h_{\mathcal{W}}^{\alpha \beta}$. The integration concepts change slightly depending on whether the field point is in the near or the wave-zone. As a method, DIRE is theory agnostic and can be adapted to any theory of gravity provided a set of field equations. The present work is dedicated to calculate the near-zone expansion in TMST.
        
        As mentioned before, STTs belong to the most studied generalizations of GR. Naturally, they have also been studied in the context of PN approximations. DIRE was already adapted to calculate a PN expansion for a wide class of single STT \cite{Mirshekari:2013vb, Lang:2013fna, Lang:2014osa, Lang:2014mra}, including an equation of motion to 2.5 PN order by Mirshekari and Will \cite{Mirshekari:2013vb} as well as an analysis of tensor gravitational waves to second PN order and a scalar waveform accurate to 1.5 PN order by Lang \cite{Lang:2013fna, Lang:2014osa, Lang:2014mra}. In addition, the metric sufficient to study light deflection at 2 PN order was examined in \cite{Xie:2007gq,Deng:2012rx}, while the generic structure of the 2 PN Lagrangian for TMST and $N$ compact bodies was derived in \cite{Damour:1995kt}. Since the standard formalism does not work for STT admitting scalarization, that can be viewed as a second order phase transition, generalizations of the PN expansions were developed in \cite{Palenzuela:2013hsa,Sennett:2016rwa, Sennett:2016klh,Khalil:2019wyy}, modeling dynamical scalarization with a resumed PN expansion and obtaining gravitational waveforms in a class of STT to 2 PN relative order. More recently, Bernard studied in a series \cite{Bernard:2018hta, Bernard:2018ivi, Bernard:2019yfz} the equations of motion in STT to 3 PN order, the resulting conserved quantities and the dipolar tidal effects. Very recently, waveforms accurate to 1.5 PN order beyond GR's standard quadrupole moment were generated \cite{Bernard:2022noq}. The PN expansion and the related gravitational waveforms up to different orders were studied also in other alternative theories of gravity, e.g. massive STT, Gauss-Bonnet gravity, and Chern-Simons theories \cite{Sotiriou:2006pq, Alsing:2011er, Yagi:2011xp, Sagunski:2017nzb, Julie:2019sab, Shiralilou:2020gah, Shiralilou:2021mfl, Battista:2021rlh}.
	    
	    The  paper is organized as follows: We briefly introduce TMST in Section \ref{sec:TMST} and rewrite the field equations coming from an Einstein frame action into a form allowing to make full use of the DIRE toolkit. Section \ref{sec:DIRE} then introduces formally the mathematics of DIRE generalized to multiple scalar fields needed in our analysis. We continue in Section \ref{sec:STRUCTURE} with the formal structure of the near-zone fields and their underlying building blocks. In there we introduce all the relevant potentials utilized in the equation of motion. Next, in Section \ref{sec:EXPANSION}, we iterate through the process of DIRE until each field is of the desired order to reach an accuracy of 2.5 PN order in the final ready-to-use equation of motion. We highlight key differences with the single scalar field theories along the way. Finally, in Section \ref{sec:emtensor}, we explain our skeletonized matter model and expand all relevant associated fields. This is followed by a brief derivation of the equation of motion in TMST and the section ends by giving the full expansion of said equation. Our results are then analyzed in the following discussion, Section \ref{sec:discussion}. We take a deeper look at the equation of motion and highlight the key structures in regards to the non-trivial target space and compact binaries.
		
	\section{TENSOR-MULTI-SCALAR THEORIES OF GRAVITY}\label{sec:TMST}
		The general form of the action in TMST has the following form \cite{Damour:1992we}
		\begin{equation}\label{eq:MSAEF}
			S :=
			\frac{1}{16 \pi G_{\star}} \int \left[ R 
			- 2 \nabla_\alpha \varphi^a \, \nabla_\beta \varphi^b \, g^{\alpha\beta} \gamma_{ab}(\varphi)
			- 4 V(\varphi) \right] \sqrt{-g} \, \mathrm{d}^4 x 
			+ S_{\text{matt}} \left[ A^2(\varphi) g_{\alpha\beta}, \Psi \right]
			\,,
		\end{equation}
		given in the conformal Einstein frame. Here,  $G_{\star}$ is the bare gravitational constant, $R$ is the Ricci scalar curvature with respect to  the Einstein frame metric $g_{\alpha\beta}$, and $V(\varphi)\ge 0$ is the potential of the scalar fields $\varphi^a$. The collective matter fields $\Psi$ are coupled to the physical Jordan frame metric ${\widetilde g}_{\alpha\beta}:= A^2(\varphi) g_{\alpha\beta}$ where the conformal factor $A(\varphi)$ is an auxiliary function to convert between the frames. The field $\varphi=(\varphi^1, \dots, \varphi^n)$ acts as generalized coordinates of the $n$-dimensional Riemannian target manifold $(T^n,\gamma_{ab})$ and collects all $n$ extra scalar fields. The signature convention for all Lorentzian metrics throughout this work is $(-,+,+,+)$. 
		
		In order to make sense of the indices crowded notation intrinsic to our subject we use the convention of Greek letters $\{\alpha,\beta,\gamma, \mu, \nu,\dots\}$ for fields with respect to the Lorentzian spacetime metric $g_{\alpha \beta}$ and the Latin letters $\{i,j,k,l,\dots\}$ for purely spatial components of said metric. The indices for the target space fields $\varphi^a$, that is with respect to the Riemannian target space metric $\gamma_{ab}$, are labeled via the different Latin letters $\{a,b,c,d,\dots\}$. By a slight abuse of notation, these last indices might be added to the left of the fields when certain functions become too crowded with labels.

		To obtain the field equations out of this action, we vary it with respect to the Einstein frame metric and the scalar fields and get
		\begin{eqnarray}
			R_{\alpha\beta} &=&
			2 \gamma_{ab}(\varphi)\nabla\!_{\alpha}\varphi^a \,\nabla\!_{\beta}\varphi^b
			+ 2 V(\varphi)g_{\alpha\beta}
			+ 8\pi G_{\star} \left(T_{\alpha\beta}
			- \frac{1}{2} T g_{\alpha\beta}\right)
			\,, \label{eq:FE_T} \\
			g^{\mu\nu} \nabla\!_{\mu} \, \nabla\!_{\nu} \varphi^a &=&
			- \gamma^a_{\ bc}(\varphi)g^{\alpha\beta}\, \nabla\!_{\alpha}\varphi^b \,\nabla\!_{\beta}\varphi^c
			+ \gamma^{ab}(\varphi)\frac{\partial V(\varphi)}{\partial \varphi^b}
			- 4\pi G_{\star} \gamma^{ab} \alpha_b(\varphi) T
		    \,. \label{eq:FE_S}
		\end{eqnarray}
		Here, the matter contribution to the scalar field equations is given in the last term of Eq. \eqref{eq:FE_S}. Making the standard assumption that the matter fields are independent of the scalar field one can easily show that (see e.g. \cite{Damour:1992we, Damour:1995kt, Horbatsch:2015bua, Doneva:2019krb, Doneva:2019ltb}) 
		\begin{equation}\label{eq:alpha_a_eq}
		    \alpha_a(\varphi) := \frac{\partial \log \left( A(\varphi) \right)}{\partial \varphi^a} \,.
		\end{equation}
		In the post-Newtonian formalism we adopt, as part of the skeletonization procedure, one can assume, however that the masses of the individual self-gravitating objects depend on the scalar fields as well \cite{Damour:1992we, Damour:1995kt, Mirshekari:2013vb}, which makes the resulting energy-momentum tensor also $\varphi$-dependent. Hence, derivatives of the trace of the energy-momentum tensor with respect to the scalar field have to be included in the field equation \eqref{eq:FE_S}. It was demonstrated in \cite{Damour:1992we, Damour:1995kt} that these derivatives can be introduced through a redefinition of $\alpha_a(\varphi)$. Here we will follow the approach of \cite{Damour:1992we}, that is to keep the expression for $\alpha_a(\varphi)$ in its general form in the first part of the analysis and only later present its explicit form when we discuss the matter fields and skeletonization.
		
		The energy-momentum tensor $T_{\alpha\beta}$ of the non-gravitational fields is defined by
		\begin{equation*}
			T_{\alpha\beta} :=
			-\frac{2}{\sqrt{-g}}\frac{\delta S_{\text{matt}}\left[A^2(\varphi)g_{\alpha\beta}, \Psi \right]}{\delta g^{\alpha \beta}}
		\end{equation*}
        The transformation of the energy momentum tensor between the two frames is given by
		\begin{equation*}
			\widetilde{T}_{\alpha\beta} \,=\, A^{-2}(\varphi)T_{\alpha\beta}\,,
		\end{equation*}
		where $\widetilde{T}_{\alpha\beta}$ is the corresponding energy-momentum tensor in the physical Jordan frame.

        Let us comment further on the frame that we will be using. Throughout the calculations we adopt the Einstein frame that is much more convenient from a mathematical point of view \cite{Fierz:1956zz}. We will refer to the physical Jordan frame when necessary. This was the approach followed by \cite{Damour:1992we,Damour:1995kt}, where the post-Newtonian formalism in TMST was explored for the first time. The analysis in the single scalar field theory up to 2.5 PN order, though, was performed in the Jordan frame \cite{Mirshekari:2013vb}. That is why even though we follow the DIRE formalism of \cite{Mirshekari:2013vb} there are some important differences related to the use of a different frame that will be discussed below. As a matter of fact the Einstein frame is very natural for the definition of TMST \cite{Damour:1992we, Damour:1995kt} because of the presence of multiple scalar fields. The freedom to choose a conformal factor $A(\varphi)$ that might depend on them in a non-trivial way and the relation of this factor to the nonminimal coupling between the scalar field and the Ricci scalar in the Jordan frame can lead to a significantly higher degree of complexity of the Jordan frame field equations compared to the single scalar field case. That is why we adopt the Einstein frame throughout our calculations.
	
	\section{Mathematical Setup of  DIRE}\label{sec:DIRE}
		
		In this section we will setup all the mathematics needed to fully make use of the powerful construction DIRE. Along the way, we emphasize key differences to GR and STT, and we explain how we adapted the method for our use in TMST. First, we define a quantity called \emph{gothic inverse metric}
		\begin{equation}\label{eq:goth}
			\mathfrak{g}^{\alpha \beta } := \sqrt{-g}  g^{\alpha \beta} \,,
		\end{equation}
		where $g:=\det\left( {g_{\alpha \beta}}\right)$ and $g^{\alpha \beta}$ is the inverse metric representation. As a multiplication with the metric determinant, the gothic inverse metric is not tensor but rather a tensor density. Notice that obtaining this density is sufficient to reconstruct the desired spacetime metric $g_{\alpha \beta}$ since $\det(\mathfrak{g}^{\alpha \beta })=\det(g_{\alpha \beta})=g$.
		Next, using this gothic metric, we define the tensor density 
		\begin{equation}\label{eq:Hdens}
			H^{\alpha \mu \beta \nu} :=
			\mathfrak{g}^{\alpha \beta} \mathfrak{g}^{\mu \nu}
			- \mathfrak{g}^{\alpha \nu} \mathfrak{g}^{\beta \mu }
			\,.
		\end{equation}
		Since the gothic metric inherits the symmetry of the spacetime metric, this new density here, as a difference of products of the gothic metric, brings its own symmetries. More specific, it is skew symmetric in the first, as well as, in the last two indices (i.e. $0=H^{(\alpha \mu)\beta \nu}=H^{\alpha \mu(\beta \nu)}$). It also has the interchange symmetry $H^{\alpha \mu \beta \nu}=H^{\beta \nu \alpha \mu}$. With that, this newly defined tensor density shares all the symmetric properties of the Riemann tensor. Using this, we can construct the useful identity
		\begin{equation}\label{eq:LLident}
			\partial_{\mu} \partial_{\nu} H^{\alpha \mu \beta \nu} = 2 (-g) G^{\alpha \beta} + 16 \pi G_\star (-g) t_{LL}^{\alpha \beta} \,,
		\end{equation}
		with the Einstein tensor $G^{\alpha \beta}$ and a new field called \emph{Landau-Lifshitz pseudotensor}
		\begin{eqnarray}\label{eq:LL_pseudo}
		    16 \pi G_\star (-g) t_{LL}^{\alpha \beta} &:=&
		    \partial_\lambda \mathfrak{g}^{\alpha \beta} \partial_\mu \mathfrak{g}^{\lambda \mu}
		    - \partial_\lambda \mathfrak{g}^{\alpha \lambda} \partial_\mu \mathfrak{g}^{\beta \mu}
		    + \frac{1}{2} g^{\alpha \beta} g_{\lambda \mu} \partial_\rho \mathfrak{g}^{\lambda \nu} \partial_\nu \mathfrak{g}^{\mu \rho}
		    \nonumber \\
            &&
            - g^{\alpha \lambda} g_{\mu \nu} \partial_\rho \mathfrak{g}^{\beta \nu} \partial_\lambda \mathfrak{g}^{\mu \rho}
            - g^{\beta \lambda} g_{\mu \nu} \partial_\rho \mathfrak{g}^{\alpha \nu} \partial_\lambda \mathfrak{g}^{\mu \rho}
            + g^{\nu \rho} g_{\lambda \mu} \partial_\nu \mathfrak{g}^{\alpha \lambda} \partial_\rho \mathfrak{g}^{\beta \mu}
            \nonumber \\
            &&
            + \frac{1}{8} \left( 2 g^{\alpha \lambda} g^{\beta \mu} - g^{\alpha \beta} g^{\lambda \mu} \right)
            \left( 2 g_{\nu \rho} g_{\sigma \tau} - g_{\rho \sigma} g_{\nu \tau} \right)
            \partial_\lambda \mathfrak{g}^{\nu \tau} \partial_\mu \mathfrak{g}^{\rho \sigma}
            \,.
		\end{eqnarray}
		Equation \eqref{eq:LLident} is really the starting point of post-Minkowskian theory which, in turn, is the foundation of post-Newtonian theory used in this paper. These field equations are as usual of second order: We see that the left-hand side has second derivatives of the tensor density $H^{\alpha \mu \beta \nu}$ which implies second derivatives of the gothic metric and hence also the spacetime metric. Notice that all these are pure geometric quantities, exactly as the left-hand side of the standard Einstein field equations. In order to put all the physical quantities on the right-hand side, we need to rewrite the first part of our derived field equations \eqref{eq:FE_T} in terms of the Einstein tensor. To do that, we collect all scalar field related terms in \eqref{eq:FE_T} via
		\begin{equation}
			\frac{16 \pi G_\star}{(-g)} t^\varphi_{\alpha\beta} := 
			- 2 g_{\alpha\beta} \gamma_{ab}(\varphi)g^{\mu\nu} \nabla\!_{\mu}\varphi^a \,\nabla\!_{\nu}\varphi^b
			+ 4 \gamma_{ab}(\varphi) \nabla\!_{\alpha}\varphi^a \,\nabla\!_{\beta}\varphi^b
			- 4V(\varphi)g_{\alpha\beta}
			\, ,
		\end{equation}
		and we rewrite \eqref{eq:FE_T} as
		\begin{equation}\label{eq:TMS_FE}
			G_{\alpha\beta} = 8\pi G_{\star}T_{\alpha\beta} + \frac{8 \pi G_\star}{(-g)} t^\varphi_{\alpha\beta} \,.
		\end{equation}
		The two equations above visualize the difference to pure GR quite directly as the additional terms are all encoded in $t^\varphi_{\alpha \beta}$. Thanks to this specific form, we can insert this Eq. \eqref{eq:TMS_FE} into our newly defined field equation \eqref{eq:LLident} and obtain
		\begin{equation}\label{eq:LLident_TMS}
			\partial_{\mu} \partial_{\nu} H^{\alpha \mu \beta \nu} =
			16 \pi G_{\star} (-g) \left( T^{\alpha\beta}
			+ t^{\alpha\beta}_{LL}
			+ t^{\alpha\beta}_{\varphi} \right)
			\, .
		\end{equation}
        Now we can take a closer look on the right-hand side. The physical quantities are as usual incorporated in the energy-momentum tensor $T^{\alpha \beta}$. As before, the difference to GR is the inclusion of the field $t^{\alpha \beta}_\varphi$. Note that while Eq. \eqref{eq:LLident_TMS} looks algebraically identical to the single scalar field case in \cite{Mirshekari:2013vb}, the difference is hidden in the definition of $t^{\alpha \beta}_\varphi$ as all the multiple scalar fields are contracted in there.
        
        Until now, all manipulations done in this section can be seen as equivalent to the theory defined by the field equation \eqref{eq:FE_T}, and we have yet to focus our attention on the extra field equations \eqref{eq:FE_S}. The next steps we take show explicitly the advantage of this Landau-Lifshitz formulation of gravity. The goal is to isolate the spacetime metric potentials as well as the multiple scalar fields in a flat wave equation. Once we succeed in doing that, we can make full use of PDE theory and solve for those desired fields.

		To proceed, we impose \emph{harmonic coordinate (gauge) conditions} $\partial_{\beta} \mathfrak{g}^{\alpha \beta} = 0$ and define the potentials
		\begin{equation}\label{eq:HGC}
			h^{\alpha \beta} := \eta^{\alpha \beta} - \mathfrak{g}^{\alpha \beta}
		\end{equation}
		with inverse Minkowski metric $\eta^{\alpha \beta}$ in Lorentzian coordinates $(t:=x^0,x^j)$. It is easily verified that such coordinates always exist as our formulation inherits the coordinate freedom from the usual GR formulation. The fields $h^{\alpha \beta}$ are, next to the scalar fields, the main focus of the rest of this paper. Together, they serve as the unknowns in the wave equations that we will derive below. First, note that knowing all $h^{\alpha \beta}$ is sufficient to recreate the spacetime metric via the definition \eqref{eq:HGC} and Eq. \eqref{eq:goth}. Next, it is easy to imagine that any ansatz to solve for $h^{\alpha \beta}$ works best if the spacetime geometry is close to flat Minkowski space since then Eq. \eqref{eq:HGC} implies that the $h^{\alpha \beta}$ are small and, in some sense, a perturbation to a flat background. The above introduced harmonic gauge translates nicely to these fields to $\partial_{\beta} h^{\alpha \beta} = 0$.
		
		Using all this, the left-hand side of \eqref{eq:LLident_TMS} becomes
		\begin{equation}\label{eq:new_H_ident}
			\partial_{\mu} \partial_{\nu} H^{\alpha \mu \beta \nu} = -\Box h^{\alpha\beta} - 16 \pi G_{\star} (-g) t^{\alpha\beta}_H \,,
		\end{equation}
		for the usual harmonic-gauge pseudotensor
		\begin{equation}
		    16 \pi G_\star (-g) t^{\alpha\beta}_H := \partial_\mu h^{\alpha \nu} \partial_\nu h^{\beta \mu} - h^{\mu \nu} \partial_\mu \partial_\nu h^{\alpha \beta}
		\end{equation}
		and for the Minkowskian wave operator $\Box:=\eta^{\mu \nu} \partial_\mu \partial_\nu$. The right-hand side of \eqref{eq:LLident_TMS} also simplifies as some terms in the Landau-Lifshitz pseudotensor \eqref{eq:LL_pseudo} vanish due to our chosen gauge. Now, substituting the identity \eqref{eq:LLident_TMS} into Eq. \eqref{eq:new_H_ident} and isolating the box operator to the left-hand side, we arrive at the flat wave equation
		\begin{equation}
			\Box h^{\alpha\beta} = -16 \pi G_{\star} \tau^{\alpha\beta}\,,
		\end{equation}
		where the source
		\begin{equation}
			\tau^{\alpha \beta} := 
			(-g) \left( T^{\alpha \beta}
			+ t^{\alpha \beta}_{LL} 
			+ t^{\alpha \beta}_H
			+ t^{\alpha \beta}_{\varphi} \right) 
		\end{equation}
		plays the role of an effective energy-momentum pseudotensor. Again, this source has, compared to GR, the additional term $t^{\alpha\beta}_{\varphi}$ encoding the multiple scalar field contributions. In addition, similar to the conservation of energy-momentum $\widetilde{\nabla}\!_\beta \widetilde{T}^{\alpha \beta}=0$ in the physical Jordan frame, our conformal effective source necessarily obeys an equivalent conservation law in
		\begin{equation}\label{eq:conlaw}
			\partial_\beta \tau^{\alpha\beta} = 0 \,.
		\end{equation}
		The important conceptual difference in those two described conservation laws is, however, that the first one is fundamental in the sense that this should be true in any viable theory and the latter is a consequence of assuming our field equations to be fulfilled.
		
		The second part of the field equations, Eq. \eqref{eq:FE_S}, already has the form of a wave equation with respect to a curved metric. We can, however, transform it to a flat wave equation via
		\begin{equation}
			g^{\mu\nu} \nabla\!_{\mu} \nabla\!_{\nu} \varphi^a
			= \frac{1}{\sqrt{-g}} \partial_\mu \left( \mathfrak{g}^{\mu\nu} \partial_\nu \varphi^a \right) 
			= \frac{1}{\sqrt{-g}} \left( \Box \varphi^a 
			- h^{\alpha\beta} \partial_\alpha \partial_\beta \varphi^a \right)
			\,.
		\end{equation} 
		Hence, Eq. \eqref{eq:FE_S} can be brought into the form
		\begin{equation}
			\Box \varphi^a = -8\pi G_{\star} \tau_\varphi^a \,,
		\end{equation}
		where the source is given as
		\begin{eqnarray}\label{eq:tau_src}
			\tau_\varphi^a &:=&
			- \frac{1}{8\pi G_{\star}} \sqrt{-g} \left[
			- \gamma^a_{bc} (\varphi) g^{\alpha\beta} \, \nabla\!_{\alpha} \varphi^b \,\nabla\!_{\beta} \varphi^c
			+ \gamma^{ab} (\varphi) \frac{\partial V(\varphi)}{\partial \varphi^b}
			- 4\pi G_{\star} \gamma^{ab} \alpha_b(\varphi) T \right]
			\nonumber \\
		    &&
		    - \frac{1}{8\pi G_{\star}} h^{\alpha\beta} \partial_\alpha \partial_\beta \varphi^a
		    \,.
		\end{eqnarray}
		The difference of TMST to STT becomes clearer in the source term \eqref{eq:tau_src}. For each scalar field $\varphi^a$, the target space metric $\gamma^{ab}$ and its associated Christoffel symbols $\gamma^{a}_{bc}$ contribute differently to the source. Hence, the curvature of the $n$-dimensional Riemannian target manifold $(T^n,\gamma_{ab})$ decides directly the difference in the source term. Also, one may introduce symmetry conditions to the metric $\gamma_{ab}$ to reduce certain degrees of freedom.
		
		In total, we have the system of wave equations 
		\begin{eqnarray}
			\Box h^{\alpha\beta} &=& -16 \pi G_{\star} \tau^{\alpha\beta}
			\,, \label{eq:wave_eq_h} \\
			\Box \varphi^a &=& -8\pi G_{\star} \tau_\varphi^a 
			\,. \label{eq:wave_eq_phi}
		\end{eqnarray}
		These equations, in the absence of any coordinate conditions, are referred to as the relaxed Einstein field equations, or, more accurately in our case, relaxed tensor-multiscalar theory field equations. The goal of this paper is to solve these two entangled partial differential equations in the near-zone. The formal solutions are given by the standard retarded Green functions
		\begin{eqnarray}
		    h^{\alpha \beta} (t, \boldsymbol{x}) &=&
		    4 \int \frac{ \tau^{\alpha \beta} \left( t-|\boldsymbol{x}-\boldsymbol{x}^\prime|, \boldsymbol{x}^\prime \right) }{ |\boldsymbol{x}-\boldsymbol{x}^\prime| } \, \mathrm{d}^3 x^\prime
		    \, , \\
		    \varphi^a (t, \boldsymbol{x}) &=&
		    2 \int \frac{ \tau^a_\varphi \left( t-|\boldsymbol{x}-\boldsymbol{x}^\prime|, \boldsymbol{x}^\prime \right) }{ |\boldsymbol{x}-\boldsymbol{x}^\prime| } \, \mathrm{d}^3 x^\prime
		    \, .
		\end{eqnarray}
		These will be calculated iteratively with the framework DIRE as explained in \cite{Pati:2002ux, poisson2014gravity}. That is, the integrals will be expanded, incorporating a slow-motion and weak-field assumption, in terms of a small parameter $\varepsilon \sim  v^2 \sim G_\star m/r$ for the characteristic mass $m$, size $r$, and velocity $v$ of the physical objects we are interested in.
		
		Now that we have defined the main equations we want to study further and solve in \eqref{eq:wave_eq_h} and \eqref{eq:wave_eq_phi}, it is worthwhile to think about the whole setting we find ourselves in. Tensor-multiscalar theory in the Einstein frame is governed by various fields. Besides the natural spacetime metric $g_{\alpha \beta}$ and the scalar fields $\varphi^a$, we have the conformal factor $A^2(\varphi)$, the target space metric $\gamma_{ab}(\varphi)$, the scalar field potential $V(\varphi)$, and, of course, the energy-momentum tensor $T_{\alpha \beta}$. The unknowns of our system \eqref{eq:wave_eq_h} and \eqref{eq:wave_eq_phi} are the tuple $(g_{\alpha \beta}, \varphi^a)$, so for a four dimensional spacetime with $n$ scalar fields that is $10+n$ total independent fields. The initial data we need to provide therefore $[A^2(\varphi), \gamma_{ab}(\varphi), V(\varphi), T_{\alpha \beta}]$ consisting of $1+n(n+1)/2+1+10=n(n+1)/2+12$ independent fields fully defining our approach.
		
		To close this section, we set ourselves up to utilize previous results. Following \cite{Pati:2002ux,Mirshekari:2013vb}, we define the quantities
		\begin{eqnarray}
			\Lambda^{\alpha \beta} &:=& 16 \pi G_{\star} (-g) \left( t_{LL}^{\alpha \beta} + t_H^{\alpha \beta}\right)
			\,, \label{eq:Lambda} \\
			\Lambda^{\alpha \beta}_\varphi &:=& 16 \pi G_{\star} (-g) t_{\varphi}^{\alpha \beta}
			\,, \label{eq:Lambdaphi}
		\end{eqnarray}
		to rewrite the metric-potential source in \eqref{eq:wave_eq_h} and obtain
		\begin{equation}\label{eq:tau}
			16 \pi G_{\star} \tau^{\alpha\beta}
			= 16 \pi G_{\star} (-g) T^{\alpha\beta}
			+ \Lambda^{\alpha \beta}
			+ \Lambda^{\alpha \beta}_\varphi
			\,.
		\end{equation}
		This equation mimics the formulas in \cite{Pati:2002ux} and \cite{Mirshekari:2013vb}. Hence, the fields defined in \eqref{eq:Lambda} and \eqref{eq:Lambdaphi} have the same algebraic structure as in GR and single STT and can be used in our analysis. 
	
		For the remaining work of this paper we restrict ourselves to a class of TMST with vanishing potential of the scalar fields $V(\varphi)=0$. Furthermore, we adapt geometric coordinate units to set the bare gravitational constant $G_\star = 1$.
	
	\section{Formal structure of the near-zone fields}\label{sec:STRUCTURE}
	    At this stage we are ready to calculate the general form of the near-zone metric. This forms the basis of the post-Newtonian theory analyzed in the presented work.
	    
		Out of convenience we assign less indices-crowded notation (as in \cite{Pati:2000vt, Mirshekari:2013vb})
		\begin{align}\label{eq:h_defs}
			N \,:=\, h^{00}
			\,, \qquad
			K^j\,:=\, h^{0j}
			\,, \qquad
			B^{ij} \,:=\, h^{ij}
			\,, \qquad
			B^{i}_{\ i} \,:=\, B
			\,.
		\end{align}
		These fields inherit their leading order in $\varepsilon$ from the energy-momentum tensor in Eq. \eqref{eq:wave_eq_h}. The exact form of the matter model we are using will be discussed in more detail in Section \ref{sec:emtensor} but for now it is sufficient to know that there exists a hierarchy of the form $T^{0i}/T^{00} \sim \sqrt{\varepsilon} \, T^{ij}/T^{00}$. Using this fact together with Eq. \eqref{eq:wave_eq_h} yields $h^{0i}/h^{00} \sim \sqrt{\varepsilon} \, h^{ij}/h^{00}$. These physical meaningful relationships can be written more handily via the shortcuts
		\begin{align}\label{eq:leading_order}
			N \,\sim \, \mathcal{O} \left( \varepsilon \right)
			\,, \qquad
			K^{j} \,\sim \, \mathcal{O} \left( \varepsilon^{3/2} \right) 
			\,, \qquad
			B^{ij} \,\sim \, \mathcal{O} \left(\varepsilon^2 \right) 
			\,, \qquad
			B \,\sim \, \mathcal{O} \left( \varepsilon^2 \right)
			\,.
		\end{align}
		We are now in a position to calculate the expansion of our Einstein frame metric in terms of the potentials \eqref{eq:h_defs} using the leading orders in \eqref{eq:leading_order}. Utilizing \eqref{eq:HGC} to get an expansion for the gothic inverse metric $\mathfrak{g}^{\alpha \beta}$ allows us to calculate the inverse spacetime metric $g^{\alpha \beta}$ via \eqref{eq:goth}. Finally, inverting this metric perturbatively yields
		\begin{subequations}
		    \label{eq:metric_expansion}
    		\begin{eqnarray}
    			g_{00} &=&
    			- \left( 1
    			- \frac{1}{2} N
    			+ \frac{3}{8} N^2
    			- \frac{5}{16} N^3 \right)
    			+ \frac{1}{2} B \left( 1
    			- \frac{1}{2} N \right)
    			+ \frac{1}{2} K^j K^j
    			+ \mathcal{O} \left( \varepsilon^4 \right)
    			\,, \\
    			g_{0i} &=&
    			- K^i \left( 1
    			- \frac{1}{2}N \right)
    			+ \mathcal{O} \left( \varepsilon^{7/2} \right)
    			\,, \\
    			g_{ij} &=&
    			\delta^{ij} \left( 1
    			+ \frac{1}{2} N
    			- \frac{1}{8} N^2 \right)
    			+ B^{ij}
    			- \frac{1}{2} B \delta^{ij}
    			+ \mathcal{O} \left( \varepsilon^3 \right)
    			\,, \\
    			(-g) &=& 1 + N - B + \mathcal{O} \left( \varepsilon^3 \right)
    			\,.
    		\end{eqnarray}
		\end{subequations}
		All covariant and contravariant components of the fields, such as $K^i$ or $B^{ij}$, are naturally interchangeable since the spatial metric to raise and lower those indices is $\delta^{ij}$. Note that these potentials do not depend explicitly on the extra scalar fields (rather via the $h^{\mu\nu}$ fields) as in the single scalar field case in \cite{Mirshekari:2013vb} since we stay in the conformal Einstein frame, whereas the cited paper converts to the physical Jordan frame at this point. If one collapses all equations down to one scalar field we can yield the relevant fields in the Einstein frame that is an auxiliary new result of this paper.
		
		Let us comment further on the form of the metric \eqref{eq:metric_expansion}. We need the metric fields to different accuracy as those potentials contribute differently to the Lagrangian responsible for the equation of motion which will be the main result of our analysis. Remembering that $\varepsilon \sim v^2$, the standard Lagrangian has the factor $(-g_{00}-2g_{0i}v^i-g_{ij}v^iv^j)^{1/2}$; hence, to obtain the equation of motion to our desired 2.5 PN order, we need $g_{00}$ to $\varepsilon^{7/2}$ but $g_{0i}$ only to $\varepsilon^{3}$ and $g_{ij}$ to $\varepsilon^{5/2}$ due to the multiplication with the velocities $v^i \sim \sqrt{\varepsilon}$ and $v^iv^j \sim \varepsilon$, respectively. Furthermore, we can analyze how and where the various fields of \eqref{eq:h_defs} enter our spacetime metric \eqref{eq:metric_expansion}. We find that the lapse $N=h^{00}$ enters all metric fields at first post-Newtonian order despite $g_{0i}$ where it enters at $\varepsilon^{5/2}$. Next, despite the obvious contribution to $g_{0i}$, the field $K^j=h^{0j}$ also enters at $\varepsilon^3$ in $g_{00}$. Hence, for the equations of motion to 1.5 PN order, $K^j$ does not contribute to the spacetime metric field $g_{00}$. Last, the spatial part $B^{ij}=h^{ij}$ enters twice only as trace, namely in $g_{00}$ and $g_{ij}$ while the full spatial part also appears in the latter. Note that the spatial part does not enter $g_{ij}$ at all for an analysis to 1.5 PN order.
		
		Following the convention in \cite{Damour:1992we} and the work built on top of it, we define underneath quantities from the energy momentum tensor $T^{\alpha\beta}$
		\begin{subequations}\label{eq:sigmas}
		    \begin{eqnarray}
    			\sigma &:=& T^{00} + T^{ii}
    			\,, \\
    			\sigma^i &:=& T^{0i}
    			\,, \\
    			\sigma^{ij} &:=& T^{ij}
    			\,, \\
    			\sigma_\varphi^a &:=& \alpha^a ( \varphi ) \,T
    			\,.
		    \end{eqnarray}
		\end{subequations}
		These densities will aid in expanding the sources of the wave equations \eqref{eq:wave_eq_h} and \eqref{eq:wave_eq_phi}. In contrast to GR and similar to STT, the field $\sigma_\varphi^a$ is added. We see it is coupled to the matter model via $\alpha^a(\varphi)$ defined in \eqref{eq:alpha_a_eq}, hence, the conformal factor $A^2(\varphi)$ makes a contribution here.
		
		As mentioned earlier, the source fields \eqref{eq:Lambda} and \eqref{eq:Lambdaphi} are algebraic equivalent with GR; hence, those fields to the required order are given in \cite{Pati:2000vt} as 
		\begin{subequations}
			\begin{eqnarray}
				\Lambda^{00} &=&
				- \frac{7}{8} (\nabla N)^2
				+ \left\{ \frac{5}{8} \dot{N}^2
				- \ddot{N} N -2 \dot{N}^{,k} K^k
				+ \frac{1}{2} K^{i,j} \left( 3 K^{j,i}
				+ K^{i,j} \right)
				\right. 
				\nonumber \\
				&&
				\left. 
				+ \dot{K}^j N^{,j}
				- B^{ij} N^{,ij}
				+ \frac{1}{4} \nabla N \cdot \nabla B
				+ \frac{7}{8} N (\nabla N)^2 \right\}
				+ \mathcal{O} (\rho \varepsilon^3)
				\,, \\
				\Lambda^{0i} &=&
				\left\{ N^{,k}( K^{k,i}
				- K^{i,k})
				+ \frac{3}{4} \dot{N} N^{,i} \right\}
				+ \mathcal{O} \left( \rho \varepsilon^{5/2} \right)
				\,, \\
				\Lambda^{ij} &=&
				\frac{1}{4} \left\{ N^{,i} N^{,j}
				- \frac{1}{2} \delta^{ij} (\nabla N)^2 \right\} 
				+ \left\{ 2 K^{k,(i} K^{j),k}
				- K^{k,i} K^{k,j}
				- K^{i,k} K^{j,k}
				+ 2N^{,(i} \dot{K}^{j)}
				+ \frac{1}{2} N^{,(i} B^{,j)}
				\right.
				\nonumber \\
				&&
				\left.
				- \frac{1}{2} N \left( N^{,i} N^{,j}
				- \frac{1}{2} \delta^{ij} (\nabla N)^2 \right)
				- \delta^{ij} \left( K^{l,k} K^{[k,l]} 
				+ N^{,k} \dot{K}^{k}
				+ \frac{3}{8} \dot{N}^2
				+ \frac{1}{4} \nabla N \cdot \nabla B \right) \right\}
				\nonumber \\
				&&
				+ \mathcal{O} (\rho \varepsilon^3)
				\,, \\
				\Lambda^{ii} &=&
				- \frac{1}{8} (\nabla N)^2 
				+ \left\{ K^{l,k} K^{[k,l]}
				- N^{,k} \dot{K}^{k}
				- \frac{1}{4}\nabla N \cdot \nabla B
				- \frac{9}{8} \dot{N}^2
				+ \frac{1}{4}N (\nabla N)^2 \right\} 
				+ \mathcal{O} (\rho \varepsilon^3)
				\,.
			\end{eqnarray}
		\end{subequations}
		To ease the reading we employ the following notation: Parentheses denote the symmetrization of a tensor with respect to those indices while square brackets denote the antisymmetrization of a tensor with respect to those indices. A comma denotes a partial derivative with respect to the spatial coordinate while a dot indicates a time derivative. Time derivatives of order three or higher will be denoted as a number in parentheses over the field. 
		
		In order to expand the extra source terms for the multiple scalar fields we rely on an asymptotic expansion in terms of $\varphi^a$ around $\varphi^a_{\infty}$, the cosmological values of the multiple scalar fields. For the target space metric $\gamma_{ab}=\gamma_{ab}(\varphi)$, this means
		\begin{equation}
		    \gamma_{ab}(\varphi) =
		    \gamma_{ab}(\varphi_{\infty})
		    + \frac{\partial \gamma_{ab}(\varphi)}{\partial \varphi^c}\bigg |_{\varphi_\infty} \left( \varphi^c
		    - \varphi^c_\infty \right)
		    + \mathcal{O} (\varphi^2)
		    \,.
		\end{equation}
		Now, in order not to overcrowd our notations, we understand every occurrence of $\gamma_{ab}$ as asymptotically evaluated, such that
		\begin{equation}
		    \gamma_{ab} \equiv \gamma_{ab}(\varphi_{\infty})
		    \,, \qquad
		    \gamma_{ab,c} \equiv \frac{\partial \gamma_{ab}(\varphi)}{\partial \varphi^c} \bigg |_{\varphi_\infty}
		    \,. 
		\end{equation}
		Using the same techniques for the Christoffel symbols $\gamma^a_{bc}=\gamma^a_{bc}(\varphi)$, we obtain
		\begin{equation}\label{eq:TS_sym1}
		    \gamma^a_{bc}(\varphi) =
		    \gamma^a_{bc}(\varphi_\infty)
		    + \frac{\partial \gamma^a_{bc}(\varphi)}{\partial \varphi^d} \bigg |_{\varphi_\infty} \left( \varphi^d
		    - \varphi^d_\infty  \right)
		    + \mathcal{O} (\varphi^2)
		    \,,
		\end{equation}
		where it is from now on again understood that
		\begin{equation}\label{eq:TS_sym2}
		    \gamma^a_{bc} \equiv \gamma^a_{bc}(\varphi_\infty)
		    \,, \qquad
		    \gamma^a_{bc,d} \equiv \frac{\partial \gamma^a_{bc}(\varphi)}{\partial \varphi^d} \bigg |_{\varphi_\infty}
		    \,.
		\end{equation}
		Without loss of generality in what follows we can assume that the cosmological value of the scalar field is zero similar to \cite{Damour:1992we}, i.e. $\varphi^a_\infty=0$. We will however, in contrast to \cite{Damour:1992we}, not make any further simplifications by choosing specific coordinates for the target space $(T^n,\gamma_{ab})$. In their analysis, field coordinates were chosen to be asymptotically geodesic; that is, the cosmological value $\varphi_\infty$ let the Christoffel symbols vanish, i.e. $\gamma^a_{bc}(\varphi_\infty) \equiv 0$. By keeping the coordinates general ourselves, we are able to identify specific spots where the geometry of the target space makes a contribution via these Christoffel symbols. The goal is to gain some insight in the physical meaning of the target space and its form.
		
		Keeping in mind that $\partial_t \sim \sqrt{\varepsilon} \, \nabla$, we can calculate the expanded scalar field source terms to be
		\begin{subequations}
		    \label{eq:Lambda_phi}
			\begin{eqnarray}
				\Lambda^{00}_\varphi &=&
				\left\{ 2 \gamma_{ab} \delta^{ij} \varphi^{a,i} \varphi^{b,j} \right\}
				+ \left\{ 4 \gamma_{ab} N \delta^{ij} \varphi^{a,i} \varphi^{b,j}
				+ 2 \gamma_{ab} \dot{\varphi}^a \dot{\varphi}^b
				+ 2 \gamma_{ab,c}  \delta^{ij} \varphi^{a,i} \varphi^{b,j} \varphi^c \right\}
				\nonumber  \\
				&&
				+ \mathcal{O} \left( \rho \varepsilon^3 \right)
				\,, \\
				\Lambda^{0i}_\varphi &=&
				- 4 \gamma_{ab} \dot{\varphi}^a \varphi^{b,i}
				+ \mathcal{O} \left( \rho \varepsilon^{5/2} \right)
				\,, \\
				\Lambda^{ij}_\varphi &=&
				2 \left\{ 4 \gamma_{ab} \varphi^{a,i} \varphi^{b,j}
				+ 2 \gamma_{ab} \delta^{ij} \delta^{kl} \varphi^{a,k} \varphi^{b,l} \right\}
				+ 2 \left\{ 2 \gamma_{ab} N \varphi^{a,i} \varphi^{b,j}
				+ \gamma_{ab} \delta^{ij} \left( N \delta^{kl} \varphi^{a,k} \varphi^{b,l}
				+ \dot{\varphi}^a \dot{\varphi}^b \right)
				\right.
				\nonumber  \\
				&&
				\left.
				2 \gamma_{ab,c} \, \varphi^{a,i} \varphi^{b,j} \varphi^c
				+ \gamma_{ab,c} \, \delta^{ij} \delta^{kl} \varphi^{a,k} \varphi^{b,l} \varphi^c \right\}
				+ \mathcal{O} \left( \rho \varepsilon^3 \right)
				\,, \\
				\Lambda^{ii}_\varphi &=&
				\left\{ 10 \gamma_{ab} \delta^{ij} \varphi^{a,i} \varphi^{b,j} \right\}
				+ 10 \left\{ \gamma_{ab} N \delta^{ij} \varphi^{a,i} \varphi^{b,j}
				+ \gamma_{ab} \dot{\varphi}^a \dot{\varphi}^b 
				+ \gamma_{ab,c} \, \delta^{ij} \varphi^{a,i} \varphi^{b,j} \varphi^c \right\}
				\nonumber  \\
				&&
				+ \mathcal{O} \left( \rho \varepsilon^3 \right)
				\,.
			\end{eqnarray}
		\end{subequations}
		The source of the scalar fields wave equations \eqref{eq:tau_src} then can be expanded as
		\begin{eqnarray}\label{eq:tau_src_expanded}
			\tau_\varphi^a &=&
			\frac{1}{2} \sigma_\varphi^a
			+ \left\{ - \frac{1}{4} N \sigma_\varphi^a
			+ \frac{1}{8\pi} \gamma^a_{bc} \delta^{ij} \varphi^{b,i} \varphi^{c,j} \right\}
			\nonumber \\
			&&
			+ \left\{ \frac{1}{8}N^2 \sigma_\varphi^a
			+ \frac{1}{8\pi} \left(
			- N \ddot{\varphi}^a
			- 2 \dot{\varphi}^{a,k} K^k
			- \varphi^{a,ij} \, B^{ij}
			- \gamma^a_{bc} \dot{\varphi}^b \dot{\varphi}^c
			+ \gamma^a_{bc,d} \, \delta^{ij} \varphi^{b,i} \varphi^{c,j} \varphi^d \right) \right\}
			\nonumber \\
			&&
			+ \mathcal{O} \left( \rho \varepsilon^3 \right)
			\,.
		\end{eqnarray}
		
        Again, we emphasize the difference to STT by examining Eqs. \eqref{eq:Lambda_phi} and \eqref{eq:tau_src_expanded} more closely. The target space metric is directly involved in contracting the scalar field indices, and hence, the geometry of the target space is directly involved here. Besides the obvious contribution of the Christoffel symbols $\gamma^a_{bc}$ in the expanded source \eqref{eq:tau_src_expanded}, one can also notice the direct contribution of the conformal factor $A(\varphi)$ in $\sigma^a_\varphi$. Actually, as the derivative of $A(\varphi)$ in direction $\varphi^a$ is contributing, and the Einstein frame in TMST gives the freedom to choose a conformal factor $A(\varphi)$, one can see that different scalar fields might behave vastly differently, depending on their dependency in $A(\varphi)$.
        
		We continue to follow the process of DIRE and give the formal near-zone expansions of the retarded Green functions. The domain of integration is a bounded time-slice, that is a spatial hypersurface $\mathcal{M}$ at a constant time $t$ bounded by a world-tube of radius $\mathcal{R}$. This radius bounds naturally the near-zone from the far-zone. We can disregard all near-zone potentials that will depend on this auxiliary cut-off  parameter $\mathcal{R}$ since they need to cancel out with their respective counterparts from the far-zone. Furthermore, the equations of the formal near-zone expansions of the retarded Green functions are again algebraically the same as in \cite{Mirshekari:2013vb} due to the inheritance of the conservation law shown in Eq. \eqref{eq:conlaw} adapted for TMST. Knowing this, the expansions to the order we need are given as
		\begin{subequations}
			\label{eq:nearzone}
			\begin{eqnarray}
				N_{\mathcal{N}} &=&
				4 \varepsilon \int_{\mathcal{M}} \frac{\tau^{00}(t,\boldsymbol{x}^\prime)}{|\boldsymbol{x}
				- \boldsymbol{x}^\prime |} \, \mathrm{d}^3  x^\prime 
				+ 2 \varepsilon^2 \partial^2_t \int_{\mathcal{M}} \tau^{00}(t,\boldsymbol{x}^\prime) |\boldsymbol{x}-\boldsymbol{x}^\prime | \, \mathrm{d}^3 x^\prime
				- \frac{2}{3} \varepsilon^{5/2} \stackrel{(3)\qquad}{{\mathcal{I}}^{kk}(t)}
				\nonumber \\
				&&
				+ \frac{1}{6} \varepsilon^3 \partial^4_t \int_{\mathcal{M}} \tau^{00}(t,\boldsymbol{x}^\prime) |\boldsymbol{x}-\boldsymbol{x}^\prime |^3 \, \mathrm{d}^3 x^\prime 
				- \frac{1}{30} \varepsilon^{7/2} \left \{ (4x^{kl}+2r^2\delta^{kl}) \stackrel{(5)\quad}{{\mathcal{I}}^{kl}(t)}
				- 4 x^k \stackrel{(5)\qquad}{{\mathcal{I}}^{kll}(t)}
				+ \stackrel{(5)\qquad}{{\mathcal{I}}^{kkll}(t)} \right\} 
				\nonumber \\
				&&
				+ N_{\partial {\mathcal{M}}}
				+ \mathcal{O} (\varepsilon^4)
				\,, \label{eq:nearzone_N} \\
				K^i_{\mathcal{N}} &=&
				4 \varepsilon^{3/2} \int_{\mathcal{M}} \frac{\tau^{0i}(t,\boldsymbol{x}^\prime)}{|\boldsymbol{x}-\boldsymbol{x}^\prime |} \, \mathrm{d}^3 x^\prime
				+ 2 \varepsilon^{5/2} \partial^2_t \int_{\mathcal{M}} \tau^{0i}(t,\boldsymbol{x}^\prime) |\boldsymbol{x}-\boldsymbol{x}^\prime | \, \mathrm{d}^3 x^\prime
				\nonumber \\
				&&
				+ \frac{2}{9} \varepsilon^3 \left \{ 3 x^k \stackrel{(4)\quad}{{\mathcal{I}}^{ik}(t)}
				- \stackrel{(4)\qquad}{{\mathcal{I}}^{ikk}(t)}
				+ 2 \varepsilon^{mik}  \stackrel{(3)\qquad}{{\mathcal{J}}^{mk}(t)} \right \}
				+ K^i_{\partial {\mathcal{M}}}
				+ \mathcal{O} \left( \varepsilon^{7/2} \right)
				\,, \label{eq:nearzone_K} \\
				B^{ij}_{\mathcal{N}} &=&
				4 \varepsilon^2 \int_{\mathcal{M}} \frac{\tau^{ij}(t,\boldsymbol{x}^\prime)}{|\boldsymbol{x}-\boldsymbol{x}^\prime |} \, \mathrm{d}^3 x^\prime 
				- 2 \varepsilon^{5/2} \stackrel{(3)\quad}{{\mathcal{I}}^{ij}(t)}
				+ 2 \varepsilon^3 \partial^2_t \int_{\mathcal{M}} \tau^{ij}(t,\boldsymbol{x}^\prime) |\boldsymbol{x}-\boldsymbol{x}^\prime | \, \mathrm{d}^3 x^\prime
				\nonumber \\
				&&
				- \frac{1}{9} \varepsilon^{7/2} \left \{ 3 r^2 \stackrel{(5)\quad}{{\mathcal{I}}^{ij}(t)}
				- 2x^k \stackrel{(5)\qquad}{{\mathcal{I}}^{ijk}(t)}
				- 8 x^k \varepsilon^{mki} \stackrel{(4)\qquad}{{\mathcal{J}}^{m|j}(t)}
				+ 6 \stackrel{(3)\qquad}{M^{ijkk}(t)} \right \}
				+ B^{ij}_{\partial {\mathcal{M}}}
				+ \mathcal{O} (\varepsilon^4)
				\,, \label{eq:nearzone_B}\\
				\varphi^a_{\mathcal{N}} &=& 
				2 \varepsilon \int_{\mathcal{M}} \frac{\tau_\varphi^a (t,\boldsymbol{x}^\prime)}{|\boldsymbol{x}-\boldsymbol{x}^\prime |} \, \mathrm{d}^3 x^\prime 
				- 2 \varepsilon^{3/2} \dot{M}_\varphi^a  
				+ \varepsilon^2 \partial^2_t \int_{\mathcal{M}} \tau_\varphi^a (t,\boldsymbol{x}^\prime) |\boldsymbol{x}-\boldsymbol{x}^\prime | \, \mathrm{d}^3 x^\prime
				\nonumber \\
				&&
				-\frac{1}{3} \varepsilon^{5/2} \left ( r^2 \stackrel{(3)\quad}{ M_\varphi^a(t)}
				- 2x^j \stackrel{(3)\quad}{^a{\mathcal{I}}^{j}_\varphi(t)}
				+ \stackrel{(3)\quad}{^a{\mathcal{I}}^{kk}_\varphi(t)} \right )
				+ \frac{1}{12} \varepsilon^3 \partial^4_t \int_{\mathcal{M}}
				\tau_\varphi^a (t,\boldsymbol{x}^\prime) |\boldsymbol{x}-\boldsymbol{x}^\prime|^3 \, \mathrm{d}^3 x^\prime 
				\nonumber \\
				&&
				- \frac{1}{60} \varepsilon^{7/2} \left \{ r^4 \stackrel{(5)\quad}{\, M_\varphi^a(t)} 
				- 4r^2 x^j \stackrel{(5)\quad}{^a{\mathcal{I}}^{j}_\varphi(t)}
				+ (4x^{kl}+2r^2\delta^{kl}) \stackrel{(5)\quad}{^a{\mathcal{I}}^{kl}_\varphi(t)}
				\right.
				\nonumber \\
				&&
				\left.
				- 4 x^k \stackrel{(5)\qquad}{^a{\mathcal{I}}^{kll}_\varphi(t)}
				+ \stackrel{(5)\qquad}{^a{\mathcal{I}}^{kkll}_\varphi(t)} \right \} 
				+ \mathcal{O} (\varepsilon^4)
				\,. \label{eq:nearzone_phi}
			\end{eqnarray}
		\end{subequations}
		
		The key difference with STT in \cite{Mirshekari:2013vb} here is, of course, the adaption of the expansion to multiple scalar fields $\varphi^a$ in the last equation. In order to present this expansion in a readable manner, we made use of the momentum already employed in \cite{Mirshekari:2013vb} and adapt it to our needs:
		\begin{subequations}
			\begin{eqnarray}
				\mathcal{I}^Q &:=&  \int_{\mathcal{M}} \tau^{00} x^Q \, \mathrm{d}^3 x
				\,, \\
				\mathcal{J}^{iQ} &:=&  \varepsilon^{ikl}\int_{\mathcal{M}} \tau^{0l} x^{kQ} \, \mathrm{d}^3 x
				\,, \\
				M^{ijQ} &:=& \int_{\mathcal{M}} \tau^{ij} x^Q \, \mathrm{d}^3 x
				\,, \\
				^a\mathcal{I}^{Q}_\varphi &:=& \int_{\mathcal{M}} \tau_\varphi^a  x^Q \, \mathrm{d}^3 x
				\,, \label{eq:Ia_int} \\
				M_\varphi^a &:=& \int_{\mathcal{M}} \tau_\varphi^a  \, \mathrm{d}^3 x
				\,.
			\end{eqnarray}
			\label{genmoment}
		\end{subequations}
		Here, $Q$ is understood as a multi-index in the following sense: Take, as an example, the scalar dipole moments $^a{\mathcal{I}}^{j}_\varphi(t)$. For those we have $Q=j$ and 
		\begin{equation}
			^a{\mathcal{I}}^{Q}_\varphi(t) = \, ^a{\mathcal{I}}^{j}_\varphi(t) = \int_{\mathcal{M}} \tau^a_\varphi x^{j} \, \mathrm{d}^3 x \,.
		\end{equation}
		As in the single scalar field scenario, the boundary terms $N_{\partial {M}}$, $K^i_{\partial \mathcal{M}}$ and $B^{ij}_{\partial \mathcal{M}}$ have no effect for the order we are interested in. They are, however, given in Appendix C in \cite{Pati:2000vt} and will have the same algebraic form for TMST.
		
		The near-zone expansions in \eqref{eq:nearzone} are essentially an expansion in terms of time derivatives and powers of $|\boldsymbol{x}-\boldsymbol{x}^\prime|$ where we integrate out $\boldsymbol{x}^\prime$ over the previously explained set $\mathcal{M}$. This expansion is suitable since for any event $(t,\boldsymbol{x})$ in the near-zone, the difference $|\boldsymbol{x}-\boldsymbol{x}^\prime|$ is small or, more precisely, $|\boldsymbol{x}-\boldsymbol{x}^\prime|<2\mathcal{R}$ for the above introduced cut-off radius $\mathcal{R}$. Examining these equations a bit more, we notice that first time derivatives are missing in, for example, the near-zone expansion of $N_\mathcal{N}=h^{00}_\mathcal{N}$. This is a direct consequence of the conservation law \eqref{eq:conlaw} together with Gauss's theorem. None such conservation law exists, however, for the extra multiple scalar fields $\varphi^a$; hence, we do have a 1.5 PN contribution term in the expansion \eqref{eq:nearzone_phi}. We deal with that fact more closely in the calculation of the potential in the next section.
		
		The potentials resulting from integrating the source terms via Eq. \eqref{eq:nearzone} will be Poisson-like in nature. We follow the notation of \cite{Pati:2000vt} further and generalize to multiscalar potentials when appropriate. We then get for any source $f$ the Poisson potential
		\begin{equation}
			P(f) :=
			\frac{1}{4\pi} \int_{\mathcal M} \frac{f(t,\boldsymbol{x}^\prime)} {|\boldsymbol{x}-\boldsymbol{x}^\prime | } \, \mathrm{d}^3 x^\prime
			\,, \quad
			\nabla^2 P(f) = -f
			\,.
		\end{equation}
		The fields stemming from the energy-matter distribution and hence the source of the wave-equations $\sigma$, $\sigma^i$, $\sigma^{ij},$ and $\sigma^a_\varphi$ inherit potentials such as
		\begin{subequations}
		    \label{eq:Sigmas}
			\begin{eqnarray}
				\Sigma (f) &:=&
				\int_{\mathcal{M}} \frac{\sigma(t,\boldsymbol{x}^\prime)f(t,\boldsymbol{x}^\prime)}{|\boldsymbol{x}-\boldsymbol{x}^\prime | } \, \mathrm{d}^3 x^\prime
				= P(4\pi\sigma f)
				\,, \\
				\Sigma^i (f) &:=&
				\int_{\mathcal{M}} \frac{\sigma^i(t,\boldsymbol{x}^\prime)f(t,\boldsymbol{x}^\prime)}{|\boldsymbol{x}-\boldsymbol{x}^\prime | } \, \mathrm{d}^3 x^\prime
				= P(4\pi\sigma^i f)
				\,, \\
				\Sigma^{ij} (f) &:=&
				\int_{\mathcal{M}} \frac{\sigma^{ij}(t,\boldsymbol{x}^\prime)f(t,\boldsymbol{x}^\prime)}{|\boldsymbol{x}-\boldsymbol{x}^\prime | } \, \mathrm{d}^3 x^\prime
				= P(4\pi\sigma^{ij} f)
				\,, \\
				\Sigma^a_\varphi (f) &:=&
				\int_{\mathcal{M}} \frac{\sigma_\varphi^a(t,\boldsymbol{x}^\prime)f(t,\boldsymbol{x}^\prime)}{|\boldsymbol{x}-\boldsymbol{x}^\prime | } \, \mathrm{d}^3 x^\prime
				= P(4\pi\sigma_\varphi^a f)
				\,,
			\end{eqnarray}
		\end{subequations}
		where we added the theory specific $\Sigma^a_\varphi$ stemming from the source $\sigma^a_\varphi$. Integrating a source against higher powers of $|\boldsymbol{x} - \boldsymbol{x}^\prime|^{-1}$, i.e. $|\boldsymbol{x} - \boldsymbol{x}^\prime|, \ |\boldsymbol{x} - \boldsymbol{x}^\prime|^3$, are commonly referred to as superpotentials \cite{Mirshekari:2013vb}. For these we introduce the notation
		\begin{subequations}
		    \label{eq:superpotentials}
			\begin{eqnarray}
				X(f) &:=&
				\int_{\mathcal{M}} {\sigma(t,\boldsymbol{x}^\prime)f(t,\boldsymbol{x}^\prime)} {|\boldsymbol{x}-\boldsymbol{x}^\prime | } \, \mathrm{d}^3 x^\prime
				\,, \\
				Y(f) &:=&
				\int_{\mathcal{M}} {\sigma(t,\boldsymbol{x}^\prime)f(t,\boldsymbol{x}^\prime)} {|\boldsymbol{x}-\boldsymbol{x}^\prime |^3 } \, \mathrm{d}^3 x^\prime
				\,,
			\end{eqnarray}
		\end{subequations}
		and likewise their counterparts and generalizations $X^i$, $X^a_\varphi$ and analogs for $Y$.
		
		To improve readability and reduce long expressions we introduce similar definitions as in \cite{Pati:2000vt} and again adapt them to our generalized formulation. The most often used potentials are the \textit{Newtonian-like} constructions
		\begin{subequations}
			\begin{eqnarray}
				U &:=&
				\int_{\mathcal{M}} \frac{\sigma(t,\boldsymbol{x}^\prime)}{|\boldsymbol{x}-\boldsymbol{x}^\prime | } \, \mathrm{d}^3 x^\prime
				= P(4\pi\sigma)
				= \Sigma(1)
				\,, \\
				U^a_\varphi &:=&
				\int_{\mathcal{M}} \frac{\sigma_\varphi^a (t,\boldsymbol{x}^\prime)}{| \boldsymbol{x}-\boldsymbol{x}^\prime |} \, \mathrm{d}^3 x^\prime
				= P \left( 4 \pi \sigma_\varphi^a \right)
				= \Sigma^a_\varphi(1)
				\,.
			\end{eqnarray}
		\end{subequations}
		
		We use the GR potentials to PN order of \cite{Pati:2000vt, Mirshekari:2013vb}:
		\begin{eqnarray}
			V^i := \Sigma^i(1)
			\,, \quad
			\Phi_1^{ij} := \Sigma^{ij}(1)
			\,, \quad
			\Phi_1 := \Sigma^{ii}(1)
		    \,, \quad
			\Phi_2 := \Sigma(U)
			\,, \quad
			X := X(1)
			\,,
		\end{eqnarray}
		and the 2 PN potentials
		\begin{eqnarray}
			&V_2^i := \Sigma^i(U)
			\,, \quad
			\varphi_2^i := \Sigma(V^i)
			\,, \qquad 
			Y := Y(1) \,,
			\,, \qquad 
			X^i := X^i(1)
			\,, \qquad
			X_1 := X^{ii}(1)
			\,,
			\nonumber \\
			&X_2 :=  X(U)
			\,, \qquad
			P_2^{ij} := P(U^{,i}U^{,j})
			\,, \qquad
			P_2 := P_2^{ii} = \Phi_2 - \frac{1}{2} U^2
			\,, \qquad
			G_1 := P (\dot{U}^2)
			\,,
			\nonumber \\
			&G_2 := P(U \ddot{U})
			\,, \qquad
			G_3 := -P(\dot{U}^{,k} V^k)
			\,, \qquad
			G_4 := P(V^{i,j}V^{j,i})
			\,, \qquad
			G_5 := -P(\dot{V}^k U^{,k})
			\,,
			\nonumber \\
			&G_6 := P(U^{,ij} \Phi_1^{ij})
			\,, \qquad
			G_7^i := P(U^{,k}V^{k,i}) + \frac{3}{4} P( U^{,i} \dot{U} )
			\,, \qquad
			H := P(U^{,ij} P_2^{ij})
			\,.
		\end{eqnarray}
		
		In order to avoid confusion with too many indices, we keep the abbreviations for potentials including target space indices to a minimum. The ones used are listed as
		\begin{eqnarray}
		    X^a_\varphi := X^a_\varphi(1)
		    \,, \qquad
		    Y^a_\varphi := Y^a_\varphi(1)
		    \,.
		\end{eqnarray}

	\section{Expansion of near-zone fields to 2.5 PN order} \label{sec:EXPANSION}
\begin{figure}
		    \centering
    		\includegraphics[]{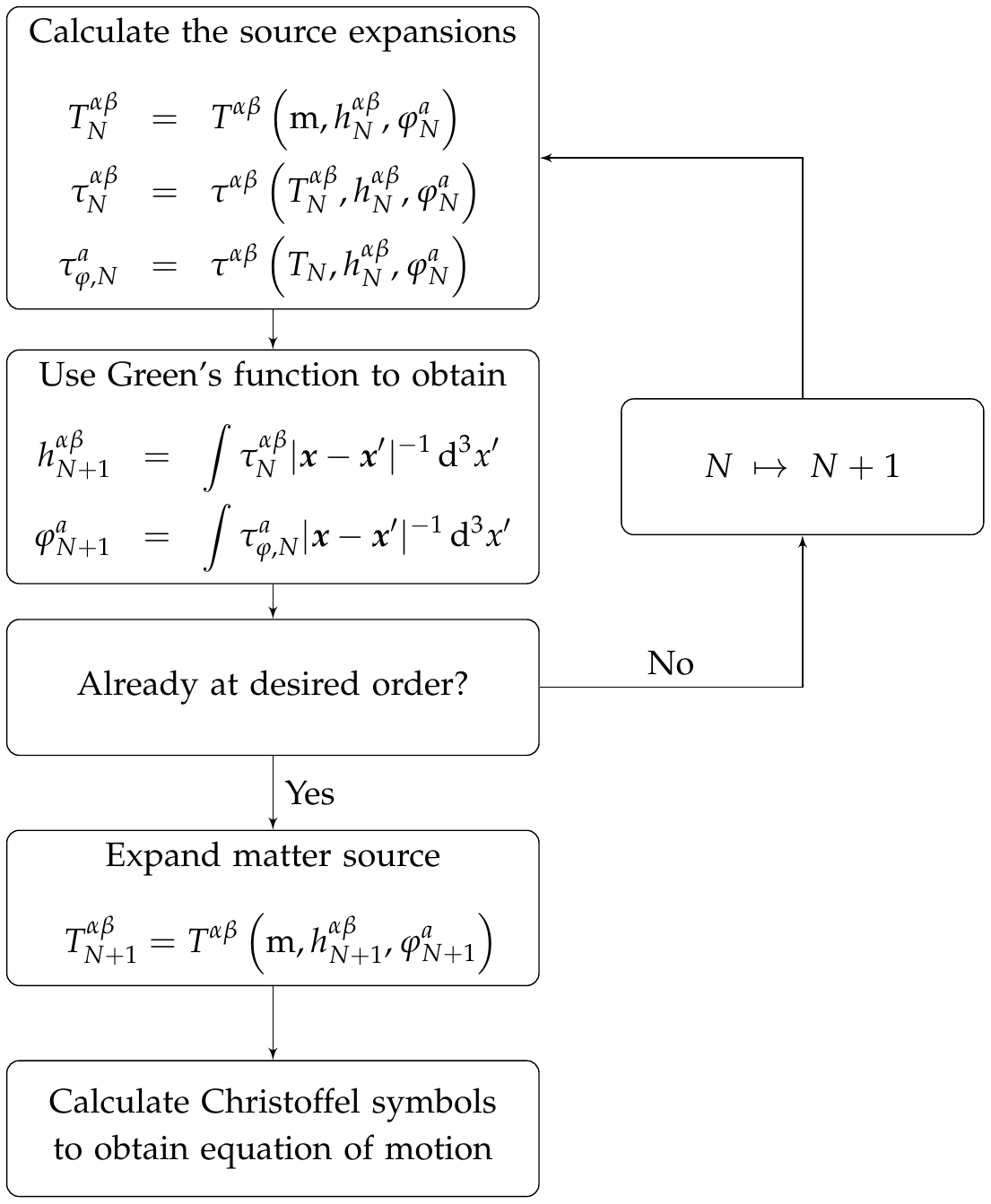}
            \caption{Flowchart of the general scheme of our calculations. This is similar to the one found in \cite{Pati:2000vt} but adapted to our TMST case here. The iterative process starts with setting $0 \equiv \varphi^a \equiv h^{\alpha \beta}$ and use that to calculate the wave equation sources in \eqref{eq:wave_eq_h}--\eqref{eq:wave_eq_phi} to lowest order. These sources are then inserted in the retarded Green's function, which in turn will be evaluated via the expansions detailed in Eqs. \eqref{eq:nearzone}. This yields the first set of the metric and scalar potentials of Eqs. \eqref{eq:field_expansions}. Now, depending on the problem of interest, one can iterate this process as long as needed, reinserting these fields in the wave equation sources and calculate those one order more accurate. Once the desired order is reached, we continue by expanding the actual matter model assumed in our work utilizing the metric and scalar field potentials calculated prior. At last, we are able to calculate the Christoffel symbols from our expanded metric which, in turn, yields the equation of motion.}
		    \label{fig:flowchart}
        \end{figure}

		We follow the convention in \cite{Mirshekari:2013vb} and \cite{Pati:2002ux} to split the metric fields in terms of their PN contributions via
		\begin{subequations}
			\label{eq:field_expansions}
			\begin{eqnarray}
				N &=&
				\varepsilon \left( N_0
				+ \varepsilon N_1
				+ \varepsilon^{3/2} N_{1.5}
				+ \varepsilon^2 N_2
				+ \varepsilon^{5/2} N_{2.5} \right)
				+ \mathcal{O} (\varepsilon^4)
				\,, \\
				K^i &=&
				\varepsilon^{3/2} \left( K_1^i
				+ \varepsilon K_2^i
				+ \varepsilon^{3/2} K_{2.5}^i \right)
				+ \mathcal{O} \left( \varepsilon^{7/2} \right)
				\,, \\
				B &=&
				\varepsilon^2 \left( B_1
				+ \varepsilon^{1/2} B_{1.5} 
				+ \varepsilon B_2
				+ \varepsilon^{3/2} B_{2.5} \right)
				+ \mathcal{O} (\varepsilon^4)
				\,, \\
				B^{ij} &=&
				\varepsilon^2 \left (B_2^{ij}
				+ \varepsilon^{1/2} B_{2.5}^{ij} \right)
				+ \mathcal{O} (\varepsilon^3)
				\,, \\
				\varphi^a &=&
				\varepsilon \left(\varphi^a_0
				+ \varepsilon^{1/2} \varphi^a_{0.5}
				+ \varepsilon \varphi^a_1
				+ \varepsilon^{3/2} \varphi^a_{1.5}
				+ \varepsilon^2 \varphi^a_2
				+ \varepsilon^{5/2} \varphi^a_{2.5} \right)
				+ \mathcal{O} (\varepsilon^4)
				\,,
			\end{eqnarray}
		\end{subequations}
		where the subscript number on each metric field denotes the leading order contribution to the equations of motion. Writing the equations in this way helps to visualize where and how much any field of interest contributes. As expected, the lapse $N=h^{00}$ and the scalar fields $\varphi^a$ are most involved as they start to contribute already at first post-Newtonian order. A complete map of the iterative process to calculate all above fields, as well as our succeeding analysis, is given in FIG. \ref{fig:flowchart}.
		
			\subsection{Calculation of Newtonian, 1 PN and 1.5 PN Fields}
			
			The lowest order in our PN expansion relies only on
			\begin{equation}
				\tau^{00} = (-g) T^{00} + \mathcal{O} (\rho\varepsilon) = \sigma + \mathcal{O} (\rho\varepsilon)
			\end{equation}
			since there are no other contributions in the source and $\sigma^{ii} \sim \varepsilon \sigma$.
			This gives us 
			\begin{equation}
				N_0 = 4 U \,.
			\end{equation}
			This result is expected since it resembles the  Newtonian  potential itself. The source to the Newtonian order of the scalar fields is given by 
			\begin{equation}
				\tau_\varphi^a = \frac{1}{2} \sigma_\varphi^a + \mathcal{O} (\rho\varepsilon)
			\end{equation}
			which returns 
			\begin{equation}\label{eq:phi_0}
				\varphi^a_0 = U^a_\varphi\,.
			\end{equation}

			To the next PN order, we substitute the field to the prior order in the source and obtain
			\begin{eqnarray}
				\tau^{00} &=&
				\sigma
				- \sigma^{ii}
				+ 4\sigma U
				- \frac{7}{8\pi} (\nabla U)^2
				+ \frac{1}{8\pi} \gamma_{ab} \delta^{ij} U_\varphi^{a,i} U_\varphi^{b,j}
				+ \mathcal{O} (\rho\varepsilon^2)
				\,, \label{eq:tau15PN} \\
				\tau^{0i} &=&
				\sigma^i + \mathcal{O} (\rho\varepsilon^{3/2})
				\,, \\
				\tau^{ii} &=&
				\sigma^{ii}
				- \frac{1}{8\pi} (\nabla U)^2
				+ \frac{5}{8 \pi} \gamma_{ab} \delta^{kl} U_\varphi^{a,k} U_\varphi^{b,l}
				+ \mathcal{O} (\rho\varepsilon^2)
				\,,  \\
				\tau^{ij} &=& \mathcal{O} (\rho\varepsilon)
				\,, \\
				\tau^a_\varphi &=&
				\frac{1}{2} \sigma_\varphi^a
				- \sigma_\varphi^a U
				+ \frac{1}{8\pi} \gamma^{a}_{bc} \delta^{ij} U_\varphi^{b,i} U_\varphi^{c,j}
				+ \mathcal{O} (\rho \varepsilon^2)
			\end{eqnarray}
			At this point it is worthwhile to compare these equations to those in \cite{Mirshekari:2013vb}. The terms presented here are natural generalizations to their single scalar field counterparts. There is, however, one notable difference: In \eqref{eq:tau15PN}, the counterpart for $\tau^{00}$ (Eq. (4.9a) in \cite{Mirshekari:2013vb}) has a term $\sigma U_s$. We do not have this term in our analysis. The reason for that is the difference of the underlying frame used.
			
			Substituting all sources above into Eqs. \eqref{eq:nearzone}, we obtain
			\begin{eqnarray}
				N_1 &=&
				7U^2 
				- 4 \Phi_1
				+ 2 \Phi_2
				+ 2 \ddot{X}
				- \gamma_{ab} \left( U^a_\varphi U^b_\varphi
				+ 2 \Sigma^a_\varphi \left( U^b_\varphi \right) \right) 
				\,, \label{eq:N_1} \\
				K_{1}^i &=& 4 V^i
				\,,\\
				B_1 &=&
				U^2
				+ 4 \Phi_1
				- 2 \Phi_2
				- 5 \gamma_{ab} \left( U^a_\varphi U^b_\varphi
				+ 2 \Sigma^a_\varphi \left( U^b_\varphi \right) \right)
				\,, \\
				\varphi^a_1 &=& 
				- \gamma^a_{bc} \left( U^b_\varphi U^c_\varphi
				+ 2 \Sigma^b_\varphi \left( U^c_\varphi \right) \right)
				- \Sigma^a_\varphi(U)
				+ \frac{1}{2} \ddot{X}^a_\varphi
				\,, \label{eq:phi_1} \\
				N_{1.5} &=&
				-\frac{2}{3} \stackrel{(3)\quad}{{\mathcal I}^{kk}(t)}
				\,,\\
				B_{1.5} &=&
				- 2 \stackrel{(3)\quad}{{\mathcal I}^{kk}(t)}
				\,, \\
				\varphi^a_{1.5} &=&
				- 2 \dot{M}^a_\varphi(t)
				+ \frac{2}{3} x^j \stackrel{(3)\quad}{^a{\mathcal I}^{j}_\varphi(t)}
				- \frac{1}{3} \stackrel{(3)\quad}{^a{\mathcal I}^{kk}_\varphi(t)}
				\,. \label{eq:phi_1.5}
			\end{eqnarray}
			
			Similar to the single scalar field case \cite{Mirshekari:2013vb} , $M^a_\varphi(t)$ is constant to the lowest PN order. This can be verified assuming that our compact bodies have a stationary internal structure together with the conservation of the baryon number in our system. Hence the term $M^a_\varphi(t)$ does not actually contribute to $\varphi^a_{0.5}$ as shown in Eq. \eqref{eq:nearzone_phi} but rather to 1.5 PN order as shown above. Hence, $\varphi^a_{0.5}$ vanishes here.
			
			\subsection{Spacetime Metric and Scalar Fields to 1.5 PN Order}
			
			In order to better see the full picture of our results until here we can put the 1 PN and 1.5 PN order contributions of Eqs. \eqref{eq:N_1}-\eqref{eq:phi_1.5} in the context of the spacetime metric \eqref{eq:metric_expansion}. This allows us to analyze the interactions between the various contributions of the expansions \eqref{eq:field_expansions} in the actual setting of the gravitational fields $g_{\alpha \beta}$. So, substituting all previous results in the 1.5 PN expansion of the metric \eqref{eq:metric_expansion} yields
			\begin{subequations}
    		    \label{eq:metric_expansion_1.5}
        		\begin{eqnarray}
        			g_{00} &=&
                    - 1 + 2 U
                    - 2 U^2
    				+ \ddot{X}
    				- 3 \gamma_{ab} \left( U^a_\varphi U^b_\varphi
    				+ 2 \Sigma^a_\varphi \left( U^b_\varphi \right) \right)
    				\nonumber \\
				    &&
				    - \frac{4}{3} \stackrel{(3)\quad}{{\mathcal I}^{kk}(t)}
        			+ \mathcal{O} \left( \varepsilon^3 \right)
        			\,, \label{eq:g00_1.5} \\
        			g_{0i} &=& - 4 V^i + \mathcal{O} \left( \varepsilon^{5/2} \right)
        			\,, \\
        			g_{ij} &=&
        			\delta^{ij} \left( 1
        			+ 2 U \right)
        			+ \mathcal{O} \left( \varepsilon^2 \right)
        			\,, \\
        			(-g) &=& 1 + 4 U + \mathcal{O} \left( \varepsilon^2 \right)
        			\,, \\
        			\varphi^a &=&
        			U^a_\varphi
        			- \gamma^a_{bc} \left( U^b_\varphi U^c_\varphi
    				+ 2 \Sigma^b_\varphi \left( U^c_\varphi \right) \right)
    				- \Sigma^a_\varphi(U)
    				+ \frac{1}{2} \ddot{X}^a_\varphi
    				\nonumber \\
				    &&
    				- 2 \dot{M}^a_\varphi(t)
    				+ \frac{2}{3} x^j \stackrel{(3)\quad}{^a{\mathcal I}^{j}_\varphi(t)}
    				- \frac{1}{3} \stackrel{(3)\quad}{^a{\mathcal I}^{kk}_\varphi(t)}
    				+ \mathcal{O} \left( \varepsilon^3 \right)
        			\,.
        		\end{eqnarray}
		    \end{subequations}
			It is worthwhile to take a step back and examine these potentials more closely. Let us start with comparing this result to the one scalar field case given in \cite{Mirshekari:2013vb}. In there, the notation is as follows: The extra scalar field given in the field equations is denoted by $\phi$ with cosmological value $\phi_0$ and the rescaling $\phi/\phi_0 =: 1-\Psi$. This gives us the opportunity to collapse our equations to one scalar field, $\varphi^a \equiv \phi$, and choose our free parameters as
			\begin{eqnarray}
			    A^2(\phi) := \frac{\phi_0}{\phi} \, ; \qquad \gamma_{ab}(\phi) \equiv \gamma_{00}(\phi) := \frac{2\omega(\phi)+3}{4\phi^2} \,.
			\end{eqnarray}
			Now, calculating the physical metric ${\widetilde g}_{\alpha\beta}$ via the conformal relation ${\widetilde g}_{\alpha\beta} = \phi_0/\phi \, g_{\alpha\beta}$, we obtain the same result as in \cite{Mirshekari:2013vb}, Eqs. (4.11).
			
			Next, let us further analyze the physical Jordan frame metric ${\widetilde g}_{\alpha\beta} = A^2(\varphi) g_{\alpha\beta}$ in our tensor-multiscalar setting. Notice that, to the here discussed 1.5 PN order, the only scalar field contribution in the tensorial part of the Einstein frame metric $g_{\alpha \beta}$ is in the $\mathrm{d}t^2$ component $g_{00}$, Eq. \eqref{eq:g00_1.5}. This is not true for the physical metric as one can see with an asymptotic expansion of the conformal factor
			\begin{equation}
			    A^2(\varphi) = A^2(\varphi_\infty) + 2 \frac{\partial A(\varphi)}{\partial \varphi^a} \bigg|_{\varphi_\infty} \left( \varphi^a - \varphi^a_\infty \right) + \mathcal{O} \left( \varphi^2 \right) \,,
			\end{equation}
			or, simply written (keeping $\varphi_\infty\equiv0$ from earlier in mind), $A^2(\varphi)=A^2_0+2A_{0,a} \varphi^a + \mathcal{O} ( \varphi^2 )$. Now, substituting in the lowest order contribution of $\varphi^a$ via Eq. \eqref{eq:phi_0}, the nontrivial contribution to the first PN order of the physical Jordan frame metric is given as
			\begin{equation}
			    \widetilde{g}_{00} = -A^2_0 + 2A^2_0 U - 2 A_{0,a} U^a_\varphi + \mathcal{O} \left( \varepsilon^{2} \right) \,.
			\end{equation}
			The form of the physical metric here makes sense as it is a linear combination of the Newtonian-like gravitational potentials $U$ and $U^a_\varphi$ while the coefficient in front can be interpreted as rescaled effective gravitational coupling constants. The fact that there is now a combination of $1+n$ contributing potentials as a post-Newtonian addition to gravity is, of course, expected as a result of coupling $n$ scalar fields to the gravitational potentials as is done in tensor-multiscalar theory studied here.
		
		\subsection{Calculation of 2 PN and 2.5 PN Fields}
		
			At 2 PN and 2.5 PN order, we obtain
			\begin{eqnarray}
				\tau^{ij} &=&
				\sigma^{ij}
				+ \frac{1}{4\pi} \left( U^{,i}U^{,j}
				- \frac{1}{2} \delta^{ij} (\nabla U)^2 \right)
				\nonumber \\
				&&
				+ \frac{1}{8\pi} \gamma_{ab} \left( 2 U_{\varphi}^{a,i} U_{\varphi}^{b,j}
				+ \delta^{ij} \delta^{kl} U_\varphi^{a,k} U_\varphi^{b,l} \right)
				+ \mathcal{O} (\rho\varepsilon^2)
				\,, \\
				\tau^{0i} &=&
				\sigma^i
				+ 4\sigma^i U
				+ \frac{2}{\pi} U^{,j} V^{[j,i]}
				+ \frac{3}{4\pi} \dot{U} U^{,i}
				- \frac{1}{4\pi} \gamma_{ab} \dot{U}^a_\varphi U_\varphi^{b,i}
				+ \mathcal{O} \left( \rho\varepsilon^{5/2} \right)
				\,.
			\end{eqnarray}
			Using Eqs. \eqref{eq:nearzone_K} and \eqref{eq:nearzone_B}, the integrals yield
			\begin{eqnarray}
				B_2^{ij} &=&
				4 \varphi_1^{ij}
				+ 4 P_2^{ij}
				- \delta^{ij} ( 2 \Phi_2 - U^2 )
				\nonumber \\ 
				&&
				+ 4 \gamma_{ab} \, ^{ab}\!P_{2\varphi}^{ij}
				- \gamma_{ab} \delta^{ij} \left( U^a_\varphi U^b_\varphi
				+ 2 \Sigma^a_\varphi \left( U^b_\varphi \right) \right)
				\,, \\
				K_2^i &=&
				8 V_2^i
				- 8 \Phi_2^i
				+ 8 U V^i
				+ 16 G_7^i
				+ 2 \ddot{X}^i
				- 4 \gamma_{ab} \delta^{ij} P \left( \dot{U}^a_\varphi U_{\varphi}^{b,j} \right)
				\,, \\
				B_{2.5}^{ij} &=&
				- 2 \stackrel{(3)\quad}{{\mathcal I}^{ij}(t)}
				\,, \\
				K_{2.5}^i &=&
				\frac{2}{3} x^k \stackrel{(4)\quad}{{\mathcal I}^{ik}(t)}
				- \frac{2}{9} \stackrel{(4)\quad}{{\mathcal I}^{ikk}(t)}
				+ \frac{4}{9} \varepsilon^{mik} \stackrel{(3)\quad}{{\mathcal J}^{mk}(t)}
				\,.
			\end{eqnarray}
			
			To calculate the source terms of our wave equations to the final order needed, we substitute all prior results of this section in Eqs. \eqref{eq:tau} and \eqref{eq:tau_src_expanded} to obtain
			\begin{eqnarray}
				\tau^{00} &=&
				\sigma
				- \sigma^{ii}
				+ 4 \sigma U
				- \frac{7}{8\pi} (\nabla U)^2
				+ \frac{1}{8\pi} \gamma_{ab} \delta^{ij}  U_\varphi^{a,i} U_\varphi^{b,j}
				\nonumber \\
				&&
				+ \sigma \left( 7U^2
				- 8 \Phi_1
				+ 2 \Phi_2
				+ 2 \ddot{X}
				- 5 \gamma_{ab} U^a_\varphi U^b_\varphi
			    - 10 \gamma_{ab} \Sigma^a_\varphi \left( U^b_\varphi \right) \right)
				- 4 \sigma^{ii} U
				\nonumber \\
				&&
				+ \frac{1}{4\pi} \left \{ \frac{5}{2} \dot{U}^2 
				- 4 U \ddot{U}
				- 8 \dot{U}^{,k} V^k
				+ 2 V^{i,j} \left( 3 V^{j,i}
				+ V^{i,j} \right)
				+ 4 \dot{V}^j U^{,j}
				- 4 U^{,ij} \Phi_1^{ij}
				\right. 
				\nonumber \\
				&& 
				\left. 
				+ 8\nabla U \cdot \nabla \Phi_1
				- 4\nabla U \cdot \nabla \Phi_2
				- \frac{7}{2} \nabla U \cdot \nabla \ddot X
				- 10U(\nabla U)^2 -4U^{,ij} \left(P_2^{ij}
				-\gamma_{ab} P \left( U^{a,i}_\varphi U^{b,j}_\varphi \right) \right)
				\right.
				\nonumber \\
				&&
				\left.
				- 6 \gamma_{ab} U^a_\varphi \nabla U \cdot \nabla U^b_\varphi 
				- 6 \gamma_{ab} \nabla U \cdot \nabla \Sigma^a_\varphi \left( U^b_\varphi \right)
				+ 4 \gamma_{ab} U \nabla U_\varphi^a \cdot \nabla U_\varphi^b
				+ \frac{1}{2} \gamma_{ab} \dot{U}^a_\varphi \dot{U}^b_\varphi
				\right.
				\nonumber \\
				&&
				\left.
				- \gamma_{ab} \gamma^b_{cd} \nabla U^a_\varphi \cdot \nabla \left( U^c_\varphi U^d_\varphi \right)
				- 2 \gamma_{ab} \gamma^b_{cd} \nabla U^a_\varphi \cdot \nabla \Sigma^c_\varphi \left( U^d_\varphi \right)
				- \gamma_{ab} \nabla U^a_\varphi \cdot \nabla \Sigma^b_\varphi(U)
				\right.
				\nonumber \\
				&&
				\left.
				+ \frac{1}{2} \gamma_{ab} \nabla U^a_\varphi \cdot \nabla \ddot{X}^b_\varphi
				+ \frac{1}{2} \gamma_{ab,c} U_\varphi^c \nabla U_\varphi^a \cdot \nabla U_\varphi^b \right \}
				\nonumber \\
				&&
				+ \frac{4}{3} \sigma \stackrel{(3)\quad}{{\mathcal I}^{kk}(t)} 
				+ \frac{1}{2\pi} U^{,ij} \stackrel{(3)\quad}{{\mathcal I}^{ij}(t)}
				+ \mathcal{O} (\rho \varepsilon^3)
				\,, \\
				\tau^{ii} &=&
				\sigma^{ii}
				- \frac{1}{8\pi} (\nabla U)^2
				+ \frac{5}{8\pi} \gamma_{ab} \delta^{ij}  U_\varphi^{a,i} U_\varphi^{b,j}
				+ 4\sigma^{ii} U
				\nonumber \\
				&&  
				- \frac{1}{4\pi} \left \{ \frac{9}{2} \dot{U}^2 
				+ 4 V^{i,j} V^{[i,j]}
				+ 4 \dot{V}^j U^{,j} 
				+ \frac{1}{2} \nabla U \cdot \nabla \ddot{X}
				- \frac{1}{2} \gamma_{ab} U^a_\varphi \nabla U \cdot \nabla U^b_\varphi
				\right.
				\nonumber \\
				&&
				\left.
				+ \frac{1}{2} \gamma_{ab} \nabla U \cdot \nabla \Sigma^a_\varphi \left( U^b_\varphi \right)
				- \frac{5}{2} \gamma_{ab} \dot{U}^a_\varphi \dot{U}^b_\varphi
				- \frac{5}{2} \gamma_{ab} \nabla U^a_\varphi \cdot \nabla \ddot{X}^b_\varphi
				\right.
				\nonumber \\
				&&
				\left.
				+ 5 \gamma_{ab} \gamma^b_{cd} \nabla U^a_\varphi \cdot \nabla \left( U^c_\varphi U^d_\varphi \right)
				+ 10 \gamma_{ab} \gamma^b_{cd} \nabla U^a_\varphi \cdot \nabla \Sigma^c_\varphi \left( U^d_\varphi \right)
				+ 5 \gamma_{ab} \nabla U^a_\varphi \cdot \nabla \Sigma^b_\varphi(U)
				\right.
				\nonumber \\
				&&
				\left.
				- 10 \gamma_{ab} U \nabla U_\varphi^{a} \cdot \nabla U_\varphi^{b}
				- \frac{5}{2} \gamma_{ab,c} U^c_{\varphi} \nabla U_\varphi^{a} \cdot \nabla U_\varphi^{b} \right \}
				+ \mathcal{O} (\rho \varepsilon^3)
				\,, \\
				\tau^a_\varphi &=&
				\frac{1}{2} \sigma_\varphi^a
				- \sigma_\varphi^a U
				+ \frac{1}{8\pi} \gamma^a_{bc} \delta^{ij} U_\varphi^{b,i} U_\varphi^{c,j}
				\nonumber \\
				&&
				+ \sigma_\varphi^a \left( \frac{21}{4} U^2
				+ \Phi_1
				- \frac{1}{2}\Phi_2
				- \frac{1}{2}\ddot{X}
				+ \frac{1}{4}\gamma_{bc}\left( U^b_\varphi U^c_\varphi
				+ 2 \Sigma^b \left( U^c_\varphi \right) \right) \right) 
				\nonumber \\
				&&
				+ \frac{1}{4\pi} \gamma^a_{bc} \left\{ - 2 \gamma^c_{de} U^d_\varphi \nabla U^b_\varphi \cdot \nabla U^e_\varphi
				- 2 \gamma^c_{de} \nabla U^b_\varphi \cdot \nabla \Sigma^d_\varphi \left( U^e_\varphi \right)
				- \nabla U^b_\varphi \cdot \nabla \Sigma^c_\varphi (U)
				+ \frac{1}{2} \nabla U^b_\varphi \cdot \nabla \ddot{X}^c_\varphi \right\}
				\nonumber \\
				&&
				+ \frac{1}{8 \pi} \left\{
				- \gamma_{bc}^a \dot{U}^b_\varphi \dot{U}^c_\varphi
				- 4 U \ddot{U}^a_\varphi 
				+ 8 V^k \dot{U}_\varphi^{a,k}
				+ U_\varphi^{a,ij} B_2^{ij}
				+ \gamma^a_{bc,d} \delta^{ij} U^{b,i}_{\varphi} U^{c,j}_{\varphi} U^d_{\varphi} \right\}
				\nonumber \\
				&&
				+ \frac{1}{6} \sigma_\varphi^a \stackrel{(3)\quad}{{\mathcal I}^{kk}(t)}
				- \frac{1}{4\pi} U_\varphi^{a,ij} \stackrel{(3)\quad}{{\mathcal I}^{ij}(t)}
				+ \mathcal{O} (\rho \varepsilon^3)
				\,.
			\end{eqnarray}

			Substituting this into Eqs. \eqref{eq:nearzone_N}, \eqref{eq:nearzone_B}, and \eqref{eq:nearzone_phi}, yields
			\begin{eqnarray}
				N_2 &=& 
				- 16 U \Phi_1
				+ 8 U \Phi_2
				+ 7 U \ddot{X}
				+ \frac{20}{3} U^3
				- 4 V^i V^i 
				- 16 \Sigma(\Phi_1)
				+ \Sigma( \ddot{X} )
				+ 8 \Sigma^i(V^i)
				\nonumber \\
				&&
				- 2 \ddot{X}_1
				+ \ddot{X}_2
				+ \frac{1}{6} \stackrel{(4)}{Y} 
				- 4G_1 - 16G_2 + 32G_3 + 24G_4 - 16G_5 - 16G_6 - 16H
				\nonumber \\
				&&
				+ 2 \gamma_{ab} \ddot{X}^a_\varphi \left( U^b_\varphi \right)
				+ 2 \gamma_{ab} P \left( \ddot{U}^a_\varphi U^b_\varphi \right)
				+ 6 \gamma_{ab} P \left( \dot{U}^a_\varphi \dot{U}^b_\varphi \right) 
				\nonumber \\
				&&
				- 32 \gamma_{ab} \Sigma\left( U^a_\varphi U^b_\varphi \right)
				- 52 \gamma_{ab} \Sigma \left( \Sigma^{a}_\varphi \left( U^{b}_{\varphi} \right) \right)
				+ 16 P \left( U^{,ij} \gamma_{ab} \, ^{ab}P_{2\varphi}^{ij} \right)
				+ 12 \gamma_{ab} U \Sigma^a_\varphi \left( U^b_\varphi \right)
				\nonumber \\
				&&
				- 38 \gamma_{ab} \Sigma^a_\varphi \left( U U^b_\varphi \right)
				+ 4 \gamma_{ab} \gamma^b_{cd} U^a_\varphi \Sigma^c_\varphi \left( U^d_\varphi \right)
				- 4 \gamma_{ab} \gamma^b_{cd} \Sigma^c_\varphi \left( U^a_\varphi U^d_\varphi \right)
				\nonumber \\
				&&
				- 4 \gamma_{ab} \gamma^b_{cd} \Sigma^a_\varphi \left( \Sigma^{c}_\varphi \left( U^{d}_{\varphi} \right) \right)
				+ 2 \gamma_{ab} U^a_\varphi \Sigma^b_\varphi ( U )
				- 2 \gamma_{ab} \Sigma^a_\varphi \left( \Sigma^b_\varphi ( U ) \right)
				- \gamma_{ab} U^a_\varphi \ddot{X}^b_\varphi
				\nonumber \\
				&&
				+ \gamma_{ab} \Sigma^a_\varphi \left( \ddot{X}^b_\varphi \right)
				+ 4 \gamma_{ab} \gamma^b_{cd} U^a_\varphi U^c_\varphi U^d_\varphi
				+ 4 \gamma_{ab} \gamma^b_{cd} P \left( U^a_\varphi \nabla U^c_\varphi \cdot \nabla U^d_\varphi \right)
				\nonumber \\
				&&
				- 8 \gamma_{ab} \gamma^b_{cd} \Sigma^d_\varphi \left( U^a_\varphi U^c_\varphi \right)
				- 4 \gamma_{ab} \gamma^b_{cd} \Sigma^a_\varphi \left( U^c_\varphi U^d_\varphi \right)
				- \gamma_{ab,c} U^a_\varphi U^b_\varphi U^c_\varphi
				\nonumber \\
				&&
				- \gamma_{ab,c} P \left( U^a_\varphi \nabla U^b_\varphi \cdot \nabla U^c_\varphi \right)
				+ 2 \gamma_{ab,c} \Sigma^c_\varphi \left( U^a_\varphi U^b_\varphi \right)
				+ \gamma_{ab,c} \Sigma^a_\varphi \left( U^b_\varphi U^c_\varphi \right)
				+ 12 \gamma_{ab} U U^a_\varphi U^b_\varphi
				\nonumber \\
				&&
				+ 28 \gamma_{ab} P \left( U \nabla U^a_\varphi \cdot \nabla U^b_\varphi \right)
				\,, \label{eq:N_2} \\
				B_2 &=&
				U \ddot{X}
				+ 4 V^i V^i
				- \Sigma ( \ddot{X} )
				- 8 \Sigma^i(V^i) 
				+ 16 \Sigma^{ii}(U)
				+ 2 \ddot{X}_1
				- \ddot{X}_2
				- 20 G_1 + 8 G_4 + 16 G_5
				\nonumber \\
				&&
				+ 10 \gamma_{ab} \ddot{X}^a_\varphi \left( U^b_\varphi \right)
				+ 10 \gamma_{ab} P \left( \ddot{U}^a_\varphi U^b_\varphi \right)
				+ 30 \gamma_{ab} P \left( \dot{U}^a_\varphi \dot{U}^b_\varphi \right)
				+ \gamma_{ab} U \Sigma^a_\varphi \left( U^b_\varphi \right)
				\nonumber \\
				&&
				- \gamma_{ab} \Sigma^a_\varphi \left( U U^b_\varphi \right)
				- \gamma_{ab} \Sigma \left( \Sigma^a_\varphi \left( U^b_\varphi \right) \right)
				- \gamma_{ab} U U^a_\varphi U^b_\varphi
				+ 39 \gamma_{ab} P \left( U \nabla U^a_\varphi \cdot U^b_\varphi \right)
				\nonumber \\
				&&
				+ 2 \gamma_{ab} \Sigma^a_\varphi \left( U U^b_\varphi \right)
				+ \gamma_{ab} \Sigma \left( U^a_\varphi U^b_\varphi \right)
				- 5 \gamma_{ab} U^a_\varphi \ddot{X}^b_\varphi
				+ 5 \gamma_{ab} \Sigma^a_\varphi \left( \ddot{X}^b_\varphi \right)
				\nonumber \\
				&&
				+ 20 \gamma_{ab} \gamma^b_{cd} U^a_\varphi U^c_\varphi U^d_\varphi
				+ 20 \gamma_{ab} \gamma^b_{cd} P \left( U^a_\varphi \nabla U^c_\varphi \cdot \nabla U^d_\varphi \right)
				- 60 \gamma_{ab} \gamma^b_{cd} \Sigma^d_\varphi \left( U^a_\varphi U^c_\varphi \right)
				\nonumber \\
				&&
				- 20 \gamma_{ab} \gamma^b_{cd} \Sigma^a_\varphi \left( U^c_\varphi U^d_\varphi \right)
				+ 20 \gamma_{ab} \gamma^b_{cd} U^a_\varphi \Sigma^c_\varphi \left( U^d_\varphi \right)
				- 20 \gamma_{ab} \gamma^b_{cd} \Sigma^a_\varphi \left( \Sigma^c_\varphi \left( U^d_\varphi \right) \right)
				\nonumber \\
				&&
				+ 10 \gamma_{ab} U^a_\varphi \Sigma^b_\varphi (U)
				- 10 \gamma_{ab} \Sigma^a_\varphi \left( U U^b_\varphi \right)
				- 10 \gamma_{ab} \Sigma^a_\varphi \left( \Sigma^b_\varphi ( U ) \right)
				\nonumber \\
				&&
				- 5 \gamma_{ab,c} U^a_\varphi U^b_\varphi U^c_\varphi
				- 5 \gamma_{ab,c} P \left( U^a_\varphi \nabla U^b_\varphi \cdot \nabla U^c_\varphi \right)
				+ 10 \gamma_{ab,c} \Sigma^c_\varphi \left( U^a_\varphi U^b_\varphi \right)
				\nonumber \\
				&&
				+ 5 \gamma_{ab,c} \Sigma^a_\varphi \left( U^b_\varphi U^c_\varphi \right)
				\,, \label{eq:B_2} \\
				\varphi^a_2 &=&
				\frac{1}{24} \stackrel{(4)}{Y^a_\varphi}
				- \ddot{X}^a_\varphi (U)
				+ \gamma^a_{bc} \ddot{X}^b_\varphi \left( U^c_\varphi \right)
				+ \frac{3}{2} \gamma^a_{bc} P \left( \ddot{U}^b_\varphi U^c_\varphi \right)
				+ \gamma^a_{bc} P \left( \dot{U}^b_\varphi \dot{U}^c_\varphi \right)
				\nonumber \\
				&&
				+ \frac{21}{2} \Sigma^a_\varphi \left( U^2 \right)
				+ 2 \Sigma^a_\varphi \left( \Phi_1 \right)
				- \Sigma^a_\varphi \left( \Phi_2 \right)
				- \Sigma^a_\varphi \left( \ddot{X} \right)
				+ \frac{1}{2} \gamma_{bc} \Sigma^a_\varphi \left( U^b_\varphi U^c_\varphi \right)
				\nonumber \\
				&&
				+ \gamma_{bc} \Sigma^a_\varphi \left( \Sigma^b_\varphi \left( U^c_\varphi \right) \right)
				+ \gamma^a_{bc} \gamma^c_{de} U^b_\varphi U^d_\varphi U^e_\varphi
				+ \gamma^a_{bc} \gamma^c_{de} P \left( U^b_\varphi \nabla U^d_\varphi \cdot \nabla U^e_\varphi \right)
				\nonumber \\
				&&
				- 2 \gamma^a_{bc} \gamma^c_{de} \Sigma^e_\varphi \left( U^b_\varphi U^d_\varphi \right)
				- \gamma^a_{bc} \gamma^c_{de} \Sigma^b_\varphi \left( U^d_\varphi U^e_\varphi \right)
				+ 2 \gamma^a_{bc} \gamma^c_{de} U^b_\varphi \Sigma^d_\varphi \left( U^e_\varphi \right)
				- 2 \gamma^a_{bc} \gamma^c_{de} \Sigma^d_\varphi \left( U^b_\varphi U^e_\varphi \right)
				\nonumber \\
				&&
				- 2 \gamma^a_{bc} \gamma^c_{de} \Sigma^b_\varphi \left( \Sigma^d_\varphi \left( U^e_\varphi \right) \right)
				+ \gamma^a_{bc} U^b_\varphi \Sigma^c_\varphi (U)
				- \gamma^a_{bc} \Sigma^c_\varphi \left( U U^b_\varphi \right)
				- \gamma^a_{bc} \Sigma^b_\varphi \left( \Sigma^c_\varphi (U) \right)
				\nonumber \\
				&&
				- \frac{1}{4} \gamma^a_{bc} U^b_\varphi \ddot{X}^c_\varphi
				+ \frac{1}{4} \gamma^a_{bc} \Sigma^b_\varphi \left( \ddot{X}^c_\varphi \right)
				- 4 P \left( U \ddot{U}^a_\varphi \right)
				+ 8 P \left( V^k \dot{U}^{a,k}_\varphi \right)
				\nonumber \\
				&&
			    + P \left( U^{a,ij}_\varphi B^{ij}_2 \right)
			    - \frac{1}{2} \gamma^a_{bc,d} U^b_\varphi U^c_\varphi U^d_\varphi
			    - \frac{1}{2} \gamma^a_{bc,d} P \left( U^d_\varphi \nabla U^b_\varphi \cdot \nabla U^c_\varphi \right)
			    \nonumber \\
				&&
				+ \gamma^a_{bc,d} \Sigma^b_\varphi \left( U^c_\varphi U^d_\varphi \right)
				+ \frac{1}{2} \gamma^a_{bc,d} \Sigma^d_\varphi \left( U^b_\varphi U^c_\varphi \right)
				\, \label{eq:phi_2} \\
				N_{2.5} &=& 
				- \frac{1}{15}(2x^{kl}+r^2\delta^{kl})\stackrel{(5)\quad}{{\mathcal I}^{kl}(t)}
				+ \frac{2}{15} x^k\stackrel{(5)\qquad}{{\mathcal I}^{kll}(t)}
				- \frac{1}{30} \stackrel{(5)\qquad}{{\mathcal I}^{kkll}(t)}
				\nonumber \\
				&&
				+ \frac{16}{3} U\stackrel{(3)\quad}{{\mathcal I}^{kk}(t)}
				- 4 X^{,kl} \stackrel{(3)\quad}{{\mathcal I}^{kl}(t)} 
				\,, \\
				B_{2.5} &=&  
				- \frac{1}{3} r^2 \stackrel{(5)\quad}{{\mathcal I}^{ii}(t)}
				+ \frac{2}{9} x^k \stackrel{(5)\qquad}{{\mathcal I}^{iik}(t)}
				+ \frac{8}{9} x^k \varepsilon^{mki} \stackrel{(4)\qquad}{{\mathcal J}^{mi}(t)}
				- \frac{2}{3} \stackrel{(3)\qquad}{M^{iikk}(t)}
				\,, \\
				\varphi^a_{2.5} &=&  
				- \frac{1}{3} r^2 \stackrel{(3)\quad}{M_\varphi^a(t)} 
				- 4 r^2 x^j \stackrel{(5)\quad}{^a{\mathcal I}^{j}_\varphi(t)}
				+ (4x^{kl}+2r^2\delta^{kl}) \stackrel{(5)\quad}{^a{\mathcal I}^{kl}_\varphi(t)}
				- 4 x^k \stackrel{(5)\qquad}{^a{\mathcal I}^{kll}_\varphi(t)}
				+ \stackrel{(5)\qquad}{^a{\mathcal I}^{kkll}_\varphi(t)}
				\nonumber \\
				&&
				+ \frac{1}{3} U^a_\varphi \stackrel{(3)\quad}{{\mathcal I}^{kk}(t)}
				- X^{a,kl}_\varphi \stackrel{(3)\quad}{{\mathcal I}^{kl}(t)} 
				\,. 
			\end{eqnarray}
	
	\section{Energy Momentum Tensor and its Expansion}
	\label{sec:emtensor}
	
	\subsection{Expansion of Mass Distribution}
	\label{sec:expmassdist}
	
	As a matter model we make use of the idea of modeling compact bodies as skeletonized point masses. This approach has already been used in \cite{Damour:1992we,Mirshekari:2013vb} and is based on the work in \cite{eardley1975observable, Will:1977zz}. The model involves using $\delta$-functions for encoding boundary conditions derived by the effects of scalar gravitational fields. The Einstein frame matter action then takes the form
	\begin{equation} \label{eq:action_matt}
		S_\text{matt} =
		- \sum_A \int m_A\left(\varphi \left(z^\mu_A\right) \right) \sqrt{-g_{\alpha \beta} \left( z^\mu_A \right)
		\mathrm{d}z^\alpha_A \, \mathrm{d}z^\beta_A }
		\,,
	\end{equation}
	where we sum over the various bodies $A$ (not to be conflicted with the conformal factor $A(\varphi)$), and $m_A=m_A(\varphi)$ denotes the Einstein frame masses of the objects corresponding to the worldlines $z^\mu_A$.
	
	In the matter action \eqref{eq:action_matt} it is taken into account that the mass of self-gravitating objects $m_A$ can depend explicitly on the scalar fields, i.e. to have compact objects such as neutron star or a black hole endowed with scalar hair. This will effectively bring an additional contribution to the scalar fields equation \eqref{eq:FE_S} connected with the derivative of the energy momentum tensor with respect to the scalar fields that can now be nonzero \cite{eardley1975observable, Mirshekari:2013vb}. It was demonstrated in \cite{Damour:1992we} that this additional contribution can be encoded in an elegant way in the expression of $\alpha_a(\varphi)$ appearing in Eq. \eqref{eq:FE_S}. More precisely, instead of employing Eq. \eqref{eq:alpha_a_eq} that is valid for non-self-gravitating objects, we can generalize the expression for $\alpha_a(\varphi)$ in the following way
	\begin{equation} \label{eq:alpha_mass}
		\alpha_a^A(\varphi)
		:= \frac{\partial \log\left(m_A(\varphi)\right)}{\partial \varphi^a}
		= m_A^{-1}(\varphi)\frac{\partial m_A(\varphi)}{\partial \varphi^a}
		\,.
	\end{equation}
	We see that $\alpha_a^A(\varphi)$ practically acts as an effective coupling function between compact object $A$ and the contribution of the scalar fields. Following \cite{Damour:1992we} one can show that
	\begin{equation}
	    \alpha_a^A(\varphi) = \alpha_a(\varphi) + \frac{\partial \log \left( \widetilde{m}_A(\varphi) \right)}{\partial \varphi^a} \,,
	\end{equation}
	where $\alpha_a(\varphi)$ is defined in Eq. \eqref{eq:alpha_a_eq} and $\widetilde{m}_A$ is the Jordan frame mass of the objects connected to the Einstein frame one via the conformal factor $m_A = A(\varphi) \widetilde{m}_A$. Clearly, for non-self-gravitating objects $\widetilde{m}_A$ is independent of the scalar field and the second term in the above equation is zero.
	
	For convenience one can define 
	\begin{equation}
		M_A(z_A) := m_A(\varphi) \frac{1}{\sqrt{-g(z_A)}} \frac{1}{\sqrt{-g_{\alpha\beta}(z_A) \, v^\alpha_A v^\beta_A}} \,,
	\end{equation}
	with the 4-velocities of the $A$th compact object 
	\begin{equation}
		u^\alpha_A = \frac{\mathrm{d} z^\alpha_A}{\mathrm{d} z^0_A} = \left(1, \frac{\mathrm{d} \boldsymbol{z}_A}{\mathrm{d} t}\right) \,.
	\end{equation}
	Put together, varying the action in Eq. \eqref{eq:action_matt} and inserting the quantities above yields the distributional Einstein frame energy momentum tensor
	\begin{align}\label{eq:EMT}
		T^{\alpha \beta} \left(t, \boldsymbol{x}\right)  = \sum_A M_A (t) u_A^\alpha u_A^\beta \delta^3 \left(\boldsymbol{x} - \boldsymbol{z}_A(t)\right) \,.
	\end{align}
	In order to get the matter quantity to desired order we need to expand the $\varphi$ dependent masses and hence the coupling function \eqref{eq:alpha_mass} around the asymptotic values of the extra scalar fields $\varphi^a_\infty$. Remember that without loss of generality we have assumed that these are zero similar to \cite{Damour:1992we}. Formally, this yields
	\begin{eqnarray}
		m_A(\varphi) &=&
		m_{A0}\left[ 1
		+ \alpha_a^{A0} \, \varphi^a
		+ \frac{1}{2} \left( \alpha_a^{A0} \alpha_a^{A0}
		+ \beta_{ab}^{A0} \right) \varphi^a \varphi^b
		\right.
		\nonumber \\
		&&
		\left.
		+ \frac{1}{6} \left( \alpha_a^{A0} \alpha_a^{A0} \alpha_a^{A0}
		+ \beta_{ab}^{A0} \alpha_a^{A0}
		+ \alpha_a^{A0} \, \beta_{ac}^{A0}
		+ \alpha_a^{A0} \, \beta_{bc}^{A0}
		+ \beta_{abc}^{A0} \right) \varphi^a \varphi^b \varphi^c \right]
		\nonumber \\
		&&
		+ \mathcal{O} \left( \varphi^4 \right)
		\,.
	\end{eqnarray}
	Here, as in \cite{Damour:1992we, Damour:1995kt}, we introduced the notation $m_{A0}=m_A(\varphi_\infty)$ and collected the covariant derivatives $D_a$ of the target space metric $\gamma_{ab}(\varphi)$ in the symmetric quantity
	\begin{equation}\label{eq:beta_equation}
	    \beta^{A}_{ab} := D_a \, D_b \, \log(m_A(\varphi)) = D_a \, _b\alpha_A \,,
	\end{equation}
	and $\beta^A_{abc}:=D_a \,\beta^{A}_{bc}$. The superscript $A0$ denotes an evaluation of the derivative at the background value $\varphi_\infty$.
	
	Now, following \cite{Mirshekari:2013vb}, we introduce the shorthand $m_A(\varphi)=:m_{A0}[1+\mathcal{S}(\alpha,\varphi)]+ \mathcal{O} (\varepsilon^4)$, where $\alpha$ collects all $\alpha_a$ fields. In order to expand the energy tensor completely we make use of the fact that in GR (see e.g. \cite{Pati:2002ux}) we have 
	\begin{equation}\label{eq:EMT_rhostar}
		T^{\alpha\beta} = \frac{\rho^*}{\sqrt{-g}} u^\alpha u^\beta  (u^0)^{-1} \,,
	\end{equation}
	where the newly introduced quantity $\rho^*$ satisfies the continuity equation 
	\begin{equation}
	    \partial \rho^*/\partial t + \nabla \cdot (\rho^* \boldsymbol{v} ) = 0 \,.
	\end{equation}
	As in the single scalar field case in \cite{Mirshekari:2013vb}, we can identify baryonic mass in the density $\rho^*$ as point masses via the delta distribution to get
	\begin{equation}
		\rho^* = \sum_A m_{A0} \, \delta^3 \left(\boldsymbol{x} - \boldsymbol{z}_A\right) \,.
	\end{equation}
	Substituting this in Eq. \eqref{eq:EMT_rhostar} then yields
	\begin{equation}\label{eq:EMT_com}
		T^{\alpha\beta} = \frac{\rho^*}{\sqrt{-g}} v^\alpha v^\beta  u^0 [1+\mathcal{S}(\alpha,\varphi)] \,,
	\end{equation}
	with the ordinary velocities $u^\alpha = u^0 v^\alpha$ and $v^\alpha := \mathrm{d}x^\alpha/\mathrm{d}t =  (1, {\boldsymbol{v}})$. The task for the rest of this section is to express all $\sigma$-densities related to the energy momentum tensor (Eqs. \eqref{eq:sigmas}) via $\rho^*$:
	\begin{subequations}
	    \begin{eqnarray}
    		\sigma &=& T^{00} + T^{ii} = \frac{\rho^*}{\sqrt{-g}} u^0 \left(1+v^2\right) \left[1+\mathcal{S}(\alpha,\varphi)\right]
    		\,, \\
    		\sigma^i &=& T^{0i} = \frac{\rho^*}{\sqrt{-g}} u^0 v^i \left[1+\mathcal{S}(\alpha,\varphi)\right]
    		\,, \\
    		\sigma^{ij} &=& T^{ij} = \frac{\rho^*}{\sqrt{-g}} u^0 v^i v^j \left[1+\mathcal{S}(\alpha,\varphi)\right]
    		\,.
	    \end{eqnarray} 
	\end{subequations}
	Note that these equations look algebraically similar to the single scalar field theory. The difference in our frame choice, the conformal Einstein frame, compared to the physical Jordan frame used in \cite{Mirshekari:2013vb} is hidden in the velocities and the contribution of the multiple scalar fields is encrypted in $\left[1+\mathcal{S}(\alpha,\varphi)\right]$.
	
	The updated density of the scalar fields can be calculated as
	\begin{equation}
	    \sigma_\varphi^a = -\frac{\rho^*}{u^0\sqrt{-g}} \left[ \alpha_A^a + \alpha_A^a \mathcal{S}(\alpha,\varphi) \right] \,.
    \end{equation}
    Before continuing to expand all those densities to the desired order, note that we can calculate $u^0$ via
    \begin{eqnarray}
        u^0 &=& \frac{1}{\sqrt{-g_{00} -2 g_{0i}v^i - g_{ij}v^iv^j}}
        \nonumber \\
        &=&
        1 + \varepsilon \left( \frac{1}{4} N_0
        + \frac{1}{2} v^2 \right)
        + \varepsilon^2 \left(
        - \frac{3}{32} N_0^2
        + \frac{1}{4} N_1
        + \frac{1}{4} B_1
        - v^i K_1^i
        - \frac{1}{8} N_0 v^2
        + \frac{3}{8} v^4 \right)
        \nonumber
        \\
        &&
        + \varepsilon^{5/2} \left( \frac{1}{4} N_{1.5} + \frac{1}{4} B_{1.5} \right)
        + \mathcal{O} (\varepsilon^3)
        \,,
    \end{eqnarray}
    and remember that
    \begin{equation}
        \frac{1}{\sqrt{-g}} =
        1 - \varepsilon \frac{1}{2} N_0
        + \varepsilon^2 \frac{1}{2} \left( - N_1
        + \frac{3}{4} N_0^2
        + B_1 \right)
        + \varepsilon^{5/2} \frac{1}{2} \left( -N_{1.5} + B_{1.5} \right)
        + \mathcal{O} (\varepsilon^3)
        \,.
    \end{equation}
    To expand all of the above $\sigma$-densities we need to insert the metric \eqref{eq:metric_expansion} and the expansion \eqref{eq:field_expansions} to get
    \begin{subequations}
        \label{eq:rhosources}
    	\begin{eqnarray}
            \sigma &=&
            \rho^* \biggl[ 1
            + \varepsilon \left( \frac{3}{2} v^2
            - U_\sigma
            + \alpha_a^{A0} U^a_{\varphi \sigma} \right )
        	+ \varepsilon^2 \left ( \frac{7}{8} v^4
        	+ v^2 U_\sigma 
        	- 4 v^j V_\sigma^j
        	- \frac{1}{4} N_1
        	+ \frac{3}{4} B_1
        	+ \frac{5}{2} U_\sigma^2
            \right.
            \nonumber \\
            &&
            \left.
            + \alpha_a^{A0} \varphi^a_1
            + \frac{1}{2} \left( \alpha_a^{A0} \alpha_b^{A0}
            + \beta_{ab}^{A0} \right) U^a_{\varphi \sigma} U^b_{\varphi \sigma}
            + \alpha_a^{A0} U^a_{\varphi \sigma} U_\sigma
            + \frac{3}{2} \alpha_a^{A0} U^a_\varphi v^2 \right)
            \nonumber \\
            &&
        	+ \varepsilon^{5/2} \left( 2 N_{1.5}
        	+ \alpha_a^{A0} \varphi^a_{1.5} \right) 
        	+ \mathcal{O} (\varepsilon^3) \biggr]
        	\,, \\
            \sigma^i &=&
            \rho^* v^i \biggl[ 1
            + \varepsilon \left ( v^2 - U_\sigma
            + \alpha_a^{A0} U^a_{\varphi \sigma} \right)
        	+ \mathcal{O} (\varepsilon^2) \biggr]
        	\,, \\
            \sigma^{ij} &=&
            \rho^* v^i v^j \biggl[ 1 + \mathcal{O} (\varepsilon) \biggr]
            \,, \\
            \sigma^{ii} &=&
            \rho^* v^2 \biggl[ 1
            + \varepsilon \left( \frac{1}{2} v^2
            - U_\sigma
            + \alpha_a^{A0} U^a_{\varphi \sigma} \right)
        	+ \mathcal{O} (\varepsilon^2) \biggr]
        	\,,
        \end{eqnarray}
    \end{subequations}
    where the subscript $\sigma$ in $U_\sigma$, $U^a_{\varphi \sigma}$, and $V^j_\sigma$ indicates definition via the $\sigma$-potentials.
    Similarly, the $\sigma$-densities stemming from all extra scalar fields then are given as
    \begin{eqnarray}
        \label{eq:rhoasource}
        \sigma_\varphi^a &=&
        \rho^* \biggl[ \alpha^a_A
        - \varepsilon \left( \frac{1}{2} \alpha^a_A v^2
        - 3 \alpha^a_A U_\sigma
        + \alpha^a_A \alpha_b^{A0} U^b_{\varphi \sigma} \right)
        + \varepsilon^2 \left(
        - \frac{1}{8} \alpha^a_A v^4
        + \frac{1}{2} \alpha^a_A U_\sigma v^2
        + 2 \alpha^a_A U_\sigma^2
        + 4 \alpha^a_A V_\sigma^i v^i
        \right. \nonumber \\
        &&
        \left.
        - \frac{3}{4} \alpha^a_A N_1
        + \frac{1}{4} \alpha^a_A B_1
        + \alpha^a_A \alpha_b^{A0} \varphi^b_1
        + \frac{1}{2} \alpha^a_A \left( \alpha_b^{A0} \alpha_c^{A0}
        + \beta_{bc}^{A0} \right) U^b_{\varphi \sigma} U^c_{\varphi \sigma}
        + \alpha^a_A \alpha_b^{A0} U^b_{\varphi \sigma} U_\sigma
        \right. \nonumber \\
        &&
        \left.
        - \frac{1}{2} \alpha^a_A \alpha_b^{A0} U^b_{\varphi \sigma} v^2 \right)
        + \varepsilon^{5/2} \alpha^a_A \alpha_b^{A0} \varphi^b_{1.5}
        + \mathcal{O} (\varepsilon^3) \biggr]
        \,.
    \end{eqnarray}
    
    With those new densities, one can express all other fields stemming from the potentials \eqref{eq:Sigmas} and \eqref{eq:superpotentials} in terms of the redefined sources \eqref{eq:rhosources} and \eqref{eq:rhoasource}. Similar to \cite{Mirshekari:2013vb}, to avoid overcrowding the notation we will use the same notation as before and redefine
    \begin{subequations}
		\begin{eqnarray}
			U &:=&
			\int_{\mathcal{M}} \frac{\rho^* (t,\boldsymbol{x}^\prime)}{|\boldsymbol{x}-\boldsymbol{x}^\prime | } \, \mathrm{d}^3 x^\prime
			\,, \label{eq:U_rho} \\ 
			U^a_\varphi &:=&
			\int_{\mathcal{M}} \frac{\alpha_A^a (t,\boldsymbol{x}^\prime) \rho^* (t,\boldsymbol{x}^\prime)}{| \boldsymbol{x}-\boldsymbol{x}^\prime |} \, \mathrm{d}^3 x^\prime
			\,, \label{eq:U_phi_alpha}
		\end{eqnarray}
	\end{subequations}
	with the analogous rewritten potentials and superpotentials $\Sigma$, $X$, and $Y$ as in \cite{Mirshekari:2013vb}.

    \subsection{Christoffel Symbols and their Expansion}
    
        In order to calculate the equation of motion we first need to calculate the expansions of the Christoffel symbols to our desired order. Due to working in the Einstein frame, our metric expansion \eqref{eq:metric_expansion} is algebraically the same as in pure GR in \cite{Pati:2002ux}. Hence, calculating the Christoffel symbols via the standard identity
        \begin{equation}
            \Gamma^\alpha_{\beta \gamma} = \frac{1}{2} g^{\alpha \lambda} \left( g_{\lambda \beta, \gamma} + g_{\lambda \gamma, \beta} - g_{\beta \gamma, \lambda} \right)
        \end{equation}
        yields
        
        \begin{subequations}\label{eq:ChristoffelSyms}
            \begin{eqnarray}
                \Gamma^0_{00} &=&
                - \varepsilon \dot{U}
                - \varepsilon^2 \left ( \frac{1}{4} ( \dot{N}_1
                + \dot{B}_1)
                - 4 U \dot{U}
                - 4 V^i U^{,i} \right) 
                - \varepsilon^{5/2} \dot{N}_{1.5}
                + \mathcal{O} (\varepsilon^3)
                \,, \\
                \Gamma^0_{0i} &=&
                - \varepsilon^{1/2} U^{,i}
                - \varepsilon^{3/2} \left( \frac{1}{4} (N_1^{,i} + B_1^{,i}) 
                - 4 U U^{,i} \right) 
                + \mathcal{O} \left( \varepsilon^{5/2} \right)
                \,,\\
                \Gamma^0_{ij} &=&
                \varepsilon \left( 4 V^{(i,j)} + \dot{U} \delta^{ij} \right) 
                + \mathcal{O} (\varepsilon^2)
                \,,\\
                \Gamma^i_{00} &=&
                - \varepsilon^{1/2} U^{,i}
                - \varepsilon^{3/2} \left( \frac{1}{4} ( N_1^{,i} + B_1^{,i} )
                + 4 \dot{V}^i
                - 8 U U^{,i} \right)
                - \varepsilon^{5/2} \left( \frac{1}{4} ( N_2^{,i} + B_2^{,i} )
                + \dot{K}_2^i
                - 2 N_1 U^{,i}
                \right.
                \nonumber \\
                &&
                \left.
                - 2U N_1^{,i} 
                - U B_1^{,i}
                - B_2^{ij} U^{,j}
                - 4 V^i \dot{U} 
                - 16 U \dot{V}^i
                + 8 V^j V^{j,i} 
                + 48 U^2 U^{,i} \right)
                \nonumber \\
                &&
                - \varepsilon^{3} \left( \frac{1}{4} \left(N_{2.5}^{,i}
                + B_{2.5}^{,i} \right)
                + \dot{K}_{2.5}^i
                - 2 N_{1.5} U^{,i} 
                - B_{2.5}^{ij} U^{,j} \right)
                + \mathcal{O} \left(\varepsilon^{7/2}\right)
                \,, \\
                \Gamma^i_{0j} &=&
                \varepsilon \left( \dot{U} \delta^{ij}
                - 4 V^{[i,j]} \right)
                + \varepsilon^2 \biggl( \frac{1}{4} (\dot{N}_1 - \dot{B}_1) \delta^{ij}
            	- K_2^{[i,j]}
            	+ \frac{1}{2} \dot{B}_2^{ij}  
            	- 4 U \dot{U} \delta^{ij} 
                \nonumber \\
                &&
                - 4 V^j U^{,i}
                + 16 U V^{[i,j]} \biggr)
                - \varepsilon^{5/2} \left( \frac{1}{2} \dot{N}_{1.5} \delta^{ij}
                + K_{2.5}^{[i,j]}
                - \frac{1}{2} \dot{B}_{2.5}^{ij} \right)
                + \mathcal{O} (\varepsilon^3)
                \,, \\
                \Gamma^i_{jk} &=&
                \varepsilon^{1/2} \left( U^{,k} \delta^{ij}
                + U^{,j} \delta^{ik}
                - U^{,i} \delta^{jk} \right)
                \nonumber \\
                &&
                +\varepsilon^{3/2} \left( \frac{1}{4} \left( N_{1}^{,k} \delta^{ij}
                + N_{1}^{,j} \delta^{ik}
                - N_{1}^{,i} \delta^{jk} \right)
                - \frac{1}{4} \left( B_{1}^{,k} \delta^{ij}
                + B_{1}^{,j} \delta^{ik}
                - B_{1}^{,i} \delta^{jk} \right)
                \right.
                \nonumber \\
                &&
                \left.
                - 4 U \left( U^{,k} \delta^{ij}
                + U^{,j} \delta^{ik}
                - U^{,i} \delta^{jk} \right)
                + \frac{1}{2} \left(B_2^{ij,k}
                + B_2^{ik,j}
                - B_2^{jk,i} \right) \right)
                + \mathcal{O} (\varepsilon^2)
                \,.
            \end{eqnarray}
        \end{subequations}
        Examining these Christoffel symbols more closely, we immediately recognize that the Newtonian-like potential $U$, Eq. \eqref{eq:U_rho}, contributes to each symbol at its lowest order. The way this potential contributes is either as its time derivative $\dot{U}$ or as the Newtonian acceleration field $U^{,j}$. This already suggests that the equation of motion calculated in the following section will have this acceleration potential as the lowest term and then post-Newtonian corrections added to it.

    \subsection{Equation of Motion to 2.5 PN Order}
    
    \subsubsection{Derivation of Equation of Motion}
    
        From the general contracted Bianchi identity applied to the field equation \eqref{eq:FE_T} we obtain
        \begin{equation}\label{eq:BiId}
            \nabla_\nu T^{\mu \nu} = \alpha_a(\varphi) T \, \nabla^\mu \varphi^a \,.
        \end{equation}
        This TMST version of the conservation law for the Einstein frame energy-momentum tensor naturally differs from its physical Jordan frame counterpart where the right-hand side vanishes. In our case here, the right-hand side incorporates self-gravitating effects in terms of the $\alpha_a(\varphi)$ coupled to the matter trace $T$ as explained below Eq. \eqref{eq:alpha_mass} in the previous section. Now, using the energy-momentum tensor as given in \eqref{eq:EMT},
        \begin{equation}
            T^{\mu \nu} =
            \frac{1}{\sqrt{-g}} \frac{1}{u^0} m_A(\varphi) u^\mu u^\nu \delta^3
            \left(\boldsymbol{x} - \boldsymbol{z}_A\right)
            \,,
        \end{equation}
        we can project both sides of Eq. \eqref{eq:BiId} via the operator $P_\mu^{ \ \beta}=\delta_\mu^{ \ \beta}+u_\mu \, u^\beta$. This projection yields a modified geodesic identity for each compact body of the form
        \begin{equation}
            u^\nu \nabla_\nu u^\beta = - \, \alpha_a(\varphi) \left[ \nabla^\beta \varphi^a + u^\beta u_\mu \nabla^\mu \varphi^a \right] \,.
        \end{equation}
        In this form we already see a derivative of a velocity on the left-hand side and we realize that the right-hand side depends on the scalar fields in two ways. Both involve only the derivatives due to the equation stemming from a contracted Bianchi identity but they differ in the sense that the latter contribution actually is a directional derivative along the velocity $u^{\mu}$. Now, replacing $\alpha_a(\varphi)$ with the appropriate mass dependent version of Eq. \eqref{eq:alpha_mass}
        \begin{equation}
            \alpha_a^A(\varphi)
            = \frac{\partial \log(m_A(\varphi))}{\partial \varphi^a}
            = \frac{1}{m_A(\varphi)} \frac{\partial m_A(\varphi)}{\partial \varphi^a}
            \,,
        \end{equation}
        and rewriting the covariant derivatives in terms of the Christoffel symbols \eqref{eq:ChristoffelSyms}, we obtain via a $3+1$ decomposition of the form
        \begin{equation} \label{eq:EoM}
        	\frac{dv^j}{dt} + \Gamma^j_{\alpha\beta} v^\alpha v^\beta
        	-\Gamma^0_{\alpha\beta} v^\alpha v^\beta v^j 
        	= - \frac{1}{m_A(\varphi)(u^0)^2}\frac{\partial m_A(\varphi)}{\partial \varphi^a}\left( \varphi^{a,j} - \dot{\varphi}^{a} v^j \right) \,.
        \end{equation}
    
    \subsubsection{Equation of Motion in terms of Metric Potentials}
    It is time to calculate the Equation of Motion to our desired order in terms of the 3-velocities $v^j \sim \sqrt{\varepsilon}$ and $v^0=v_0\sim \mathcal{O} (1)$. We collect all previously calculated terms and sort them according to their post-Newtonian contribution via the expansion
	\begin{equation}
		\frac{dv^j}{dt} =
		a_N^j + \varepsilon a_{PN}^j
		+ \varepsilon^{3/2} a_{1.5PN}^j
		+ \varepsilon^{2} a_{2PN}^j + \varepsilon^{5/2} a_{2.5PN}^j
		+ \mathcal{O} \left( \varepsilon^3 \right)
		\,.
	\end{equation}
	Now, substituting all relevant fields in \eqref{eq:EoM}, we obtain our final coefficients as
	\begin{equation}
	    a^j_{N} = -\alpha_a^{A0} U_\varphi^{a,j} + v_0^2 \, U^{,j}
	    \label{eq:aj_0}
	\end{equation}
	\begin{eqnarray}
	    a^j_{PN} &=&
	    4 v_0^2 \dot{V}^j
        - v_0^2 \dot{U} v^j 
        + \alpha_a^{A0} \dot{U}_\varphi^a v^j 
        - 2 v_0 v^i v^j U^{,i}
        \nonumber \\
        &&
        - \delta^{jk} v^i v^k U^{,i}
        - 4 v_0 v^i V^{j,i}
        + \frac{1}{4} v_0^2 B_1^{,j}
        + \frac{1}{4} v_0^2 N_1^{,j}
        + 2 v^2 \alpha_a^{A0} U^{a,j}_\varphi
        \nonumber \\ 
        &&
        + 2 \alpha_a^{A0} U U_\varphi^{a,j}
        - \beta_{ab}^{A0} U_\varphi^b U_\varphi^{a,j}
        - \alpha_a^{A0} \varphi_1^{a,j}
        - 8 v_0^2 U U^{,j}
        + \delta^{ik} v^i v^k U^{,j}
        \nonumber \\
        &&
        + 4 v_0 v^i V^{i,j}
        - \delta^{ij} v^i \left( 2 v_0 \dot{U}
        + v^k U^{,k} \right)
        \label{eq:aj_1}
	\end{eqnarray}
	
	\begin{equation}
	    a^j_{1.5PN} = -\alpha_a^{A0} \varphi_{1.5}^{a,j}
	    \label{eq:aj_1.5}
	\end{equation}
	
	\begin{eqnarray}
	    a^j_{2PN} &=&
	    v_0^2 \dot{K}_2^{,j}
        - 16 v_0^2 \dot{V}^{,j} U
        - v_0 \dot{B}_2^{ij} v^i 
        - \frac{1}{2} v_0 \dot{B}_1 \delta^{ij} v^i
        - \frac{1}{2} v_0 \dot{N}_1 \delta^{ij} v^i
        + 8 v_0 \dot{U} \delta^{ij} U v^i
        \nonumber \\
        &&
        - \frac{1}{4} v_0^2 \dot{B}_1 v^j
        - \frac{1}{4} v_0^2 \dot{N}_1 v^j
        - 4 \alpha_a^{A0} \left( \frac{1}{2} v^2
        + U \right) \dot{U}_\varphi^a v^j
        + \alpha_a^{A0} \dot{\varphi}_1^a v^j
        \nonumber \\
        &&
        - \alpha_a^{A0} \alpha_b^{A0} \dot{U}_\varphi^a U_\varphi^b v^j
        + \left( \alpha_a^{A0} \alpha_b^{A0}
        + \beta_{ab}^{A0} \right) \dot{U}_\varphi^a U_\varphi^b v^j 
        + 4 v_0^2 \dot{U} U v^j
        + \dot{U} \delta^{ik} v^i v^j v^k
        - 4 v_0^2 \dot{U} V^j
        \nonumber \\
        &&
        - \frac{1}{2} v_0 v^i v^j B_1^{,i}
        + \frac{1}{4} \delta^{jk} v^i v^k B_1^{,i}
        - \frac{1}{2} v^i v^k B_2^{jk,i}
        - v_0 v^i K_2^{j,i}
        - \frac{1}{2} v_0 v^i v^j N_1^{,i}
        - \frac{1}{4} \delta^{jk} v^i v^k N_1^{,i}
        \nonumber \\
        &&
        + 8 v_0 U v^i v^j U^{,i}
        + 4 \delta^{jk} U v^i v^k U^{,i}
        + 16 v_0 U v^i V^{j,i}
        + 2 v^i v^j v^k V^{k,i}
        - v_0^2 U B_1^{,j}
        - \frac{1}{4} \delta^{ik} v^i v^k B_1^{,j}
        \nonumber \\
        &&
        + \frac{1}{4} v_0^2 B_2^{,j}
        + \frac{1}{2} v^i v^k B_2^{ik,j}
        + v_0 v^i K_2^{i,j}
        - 2 v_0^2 U N_1^{,j}
        + \frac{1}{4} \delta^{ik} v^i v^k N_1^{,j}
        + \frac{1}{4} v_0^2 N_2^{,j}
        \nonumber \\
        &&
        - \alpha_a^{A0} \left( \frac{1}{2} v^2
        + U \right)^2 U_\varphi^{a,j}
        + \alpha_a^{A0} \alpha_b^{A0} \varphi_1^a U_\varphi^{b,j}
        - 2 \alpha_a^{A0} \alpha_b^{A0} \left( \frac{1}{2} v^2
        + U \right) U_\varphi^a U_\varphi^{b,j}
        \nonumber \\
        &&
        - \alpha_a^{A0} \alpha_b^{A0} \alpha_c^{A0} U_\varphi^a U_\varphi^b U_\varphi^{c,j}
        + 2 \left( \alpha_a^{A0} \alpha_b^{A0}
        + \beta_{ab}^{A0} \right) \left( \frac{1}{2} v^2
        + U \right) U_\varphi^b U_\varphi^{a,j}
        \nonumber \\
        &&
        + \frac{3}{2} \alpha_c^{A0} \left( \alpha_a^{A0} \alpha_b^{A0}
        + \beta_{ab}^{A0} \right) U_\varphi^a U_\varphi^b U_\varphi^{c,j}
        - \frac{1}{2} \left( \alpha_a^{A0} \alpha_b^{A0} \alpha_c^{A0}
        + \alpha_c^{A0} \beta_{ab}^{A0}
        + \alpha_b^{A0} \beta_{ac}^{A0}
        + \beta_{abc}^{A0} \right) U_\varphi^b U_\varphi^c U_\varphi^{a,j}
        \nonumber \\
        &&
        - \left(\alpha_a^{A0} \alpha_a^{A0}
        + \beta_{ab}^{A0} \right) \varphi_1^b U_\varphi^{a,j}
        + 2 \alpha_a^{A0} \left( \frac{5}{8} v^4
        + \frac{1}{4} B_1
        + v^2 U
        - \frac{1}{2} U^2
        + \frac{1}{4} N_1
        - K_1^i v^i \right) U_\varphi^{a,j}
        \nonumber \\
        &&
        + 2 \alpha_a^{A0} \left( \frac{1}{2} v^2
        + U \right) \varphi_1^{a,j}
        + \alpha_a^{A0} \alpha_b^{A0} U_\varphi^a \varphi_1^{b,j}
        + \left( - \alpha_a^{A0} \alpha_b^{A0}
        - \beta_{ab}^{A0} \right) U_\varphi^b \varphi_1^{a,j}
        - \alpha_a^{A0} \varphi_2^{a,j}
        \nonumber \\
        &&
        - 2 v_0^2 N_1 U^{,j}
        + 48 v_0^2 U^2 U^{,j}
        - 4 \delta^{ik} U v^i v^k U^{,j}
        + 8 v_0 v^i V^i U^{,j}
        - 16 v_0 U v^i V^{i,j}
        + 8 v_0^2 V^l V^{l,j}
        \nonumber \\
        &&
        + \frac{1}{4} \delta^{ij} v^i v^k B_1^{,k}
        - \frac{1}{2} v^i v^k B_2^{ij,k}
        - \frac{1}{4} \delta^{ij} v^i v^k N_1^{,k}
        + 4 \delta^{ij} U v^i v^k U^{,k}
        + 2  v^i v^j v^k V^{i,k}
        - v_0^2 B_2^{jl} U^{,l}
        \label{eq:aj_2}
	\end{eqnarray}
	
	\begin{eqnarray}
	    a^j_{2.5PN} &=&
	    v_0^2 \dot{K}_{2.5}^{j}
    	- v_0 \dot{B}_{2.5}^{ij} v^i
    	+ v_0 \dot{N}_{1.5} \delta^{ij} v^i
    	- \dot{N}_{1.5} v_0^2 v^j
    	+ \alpha_a^{A0} \dot{\varphi}_{1.5}^a v^j
    	- v_0 v^i K_{2.5}^{j,i}
    	+ \frac{1}{4} v_0^2 B_{2.5}^{,j}
    	\nonumber \\
        &&
    	+ v_0 v^i K_{2.5}^{i,j}
    	+ \frac{1}{4} v_0^2 N_{2.5}^{,j}
    	+ 2 \alpha_a^{A0} N_{1.5} U_\varphi^{a,j}
    	+ \alpha_a^{A0} \alpha_b^{A0} \varphi_{1.5}^a U_\varphi^{b,j}
    	\nonumber \\
        &&
    	- \left( \alpha_a^{A0} \alpha_b^{A0}
    	+ \beta_{ab}^{A0} \right) \varphi_{1.5}^b U_\varphi^{a,j}
    	+ 2 \alpha_a^{A0} \left( \frac{1}{2} v^2
    	+ U \right) \varphi_{1.5}^{a,j}
    	+ \alpha_a^{A0} \alpha_b^{A0} U_\varphi^a \varphi_{1.5}^{b,j}
    	\nonumber \\
        &&
    	- \left( \alpha_a^{A0} \alpha_b^{A0}
    	+ \beta_{ab}^{A0} \right) U_\varphi^b \varphi_{1.5}^{a,j}
    	- 2 v_0^2 N_{1.5} U^{,j}
    	- v_0^2 B_{2.5}^{jl} U^{,l}
    	\label{eq:aj_2.5}
	\end{eqnarray}
	
	\section{Discussion}\label{sec:discussion}
        In the present paper we have derived a ready to use version of the equation of motion to 2.5 post-Newtonian order in a general class of tensor-multi-scalar theories (TMST) as introduced in \cite{Damour:1992we}. To achieve this, we adapted the direct integration of the relaxed field equations approach \cite{Will:1999dq, Pati:2000vt, Pati:2002ux, Will:2005sn, Wang:2007ntb, Mitchell:2007ea} beyond general relativity and the single scalar field case. Due to the specifics of the TMST and the great simplification of the field equations, we have performed our calculations in the conformal Einstein frame similar to \cite{Damour:1992we, Damour:1995kt} and in contrast to previous PN studies in the single scalar field case that employ the physical Jordan frame \cite{Mirshekari:2013vb, Lang:2013fna, Lang:2013fna, Bernard:2018hta, Bernard:2018ivi, Bernard:2019yfz, Bernard:2022noq}. Thus, as a complementary result of our studies, the Einstein frame 2.5 PN single scalar field equation of motion follows from our results when the multiple scalar fields are contracted to one scalar field. 
        
        We have consistently performed PN expansion of the metric and the scalar field up to 2.5 order. Using a skeletonization procedure to describe matter and the compact objects in general, we have derived the generalized Binachi identity and the equation of motion in TMST. We have taken into account the possibility that the mass of a compact object can depend on the scalar field for self-gravitating objects. In all these calculations we have kept a general form of TMST admitting an arbitrary number of scalar fields and without imposing restrictions on the target space metric. 
        
        Below we will summarize some of the main differences in comparison with previous studies in the single scalar field case and in GR. We will also put emphasis on the physical interpretation of our result especially with respect to inspiraling binary compact objects.
        
	    \subsection{Target Space Involvement}
	        Among the most important differences of tensor-multiscalar theories to other alternative theories is the addition of the target space $(T^n,\gamma_{ab})$. Remember that this $n$-dimensional Riemannian manifold allows us to interpret the $n$ extra scalar fields $\varphi = (\varphi^1, \dots, \varphi^n)$ of our analyzed theory of gravity as generalized coordinates of this target space $\varphi \colon spacetime \rightarrow target \; space$ such that
	        \begin{equation}
	            \mathrm{d} \sigma^2 = \gamma_{ab}(\varphi) \, \mathrm{d} \varphi^a \, \mathrm{d} \varphi^b
	        \end{equation}
	        is the line element of said target space. The addition of this construct naturally gives rise to the physical interpretation of said space. In particular, it is of interest how the curvature of this manifold contributes to the analysis of TMST. We try to hint at some answers here. First note that we kept our work as general as possible with respect to the target space, i.e. we did not choose any coordinates or made any topology/symmetry assumptions. Those two methods are generally the first steps to simplify any equations involving the manifold $(T^n,\gamma_{ab})$. Choosing specific coordinates would allow, for example, the Christoffel symbols in Eqs. \eqref{eq:TS_sym1}-\eqref{eq:TS_sym2} to vanish at lowest order. However, since derivatives of these Christoffel symbols also enter our equations and as any choice of coordinates cannot guarantee globally both, the symbols and its derivatives, to vanish, some curvature terms will inevitably contribute to our post-Newtonian analysis here.
	        
	        First, as explained in Section \ref{sec:STRUCTURE}, the target space Christoffel symbols are contracted with the scalar fields and their derivatives in the source \eqref{eq:tau_src_expanded}. This source then is integrated in the scalar field expansion \eqref{eq:nearzone_phi}. Due to the coupling with the scalar fields, these curvature terms do not contribute to Newtonian order but rather start at first PN order in $\varphi_1^a$ as evident from Eq. \eqref{eq:phi_1}. As seen in the 1.5 PN spacetime metric \eqref{eq:metric_expansion_1.5}, $\varphi_1^a$ and hence the Christoffel symbols do not enter the gravitational fields $g_{\alpha \beta}$. This means that the tensorial waveform calculated in future work will be unaffected by these symbols to 1.5 PN order making the explicit curvature a 2 PN order effect. After that, we see multiple contributions from the Christoffel symbols in $N_2$ and $B_2$ given by Eqs. \eqref{eq:N_2}-\eqref{eq:B_2}, and, of course, $\varphi^a_2$ in Eq. \eqref{eq:phi_2}. These contributions are always linked to some form of the Newtonian-like potentials $U$ and $U^a_\varphi$ through Eqs. \eqref{eq:U_rho} and \eqref{eq:U_phi_alpha}. In $\varphi^a_2$, we even have the occurrence of products of Christoffel symbols as seen in Eq. \eqref{eq:phi_2}, making the contribution of the target space curvature even more prominent.
	        
	        From all the contributions mentioned above it is clear that the geometry of the target space has a physical relevance in the post-Newtonian motion of compact objects. While these contributions are small in the sense that they occur at a higher post-Newtonian order than, for example, the contribution of the scalar fields themselves, the number of terms containing target space curvature fields is quite numerous in $N_2$ and $B_2$ (see Eqs. \eqref{eq:N_2}-\eqref{eq:B_2}). Hence, we expect noticeable contributions from them to the tensorial waveform and therefore detectable differences from general relativity and single scalar-tensor theories.
	        
        \subsection{Self-Gravitating Bodies}
            Binaries consisting of strongly self-gravitating bodies play an important role in the study of generalized theories of gravity. In the series \cite{eardley1975observable, Will:1977zz, Will:1977wq}, Eardley and Will showed that in a wide class of Brans-Dicke theories, such binaries are governed by dipole-radiation term. This term promised some new physics as no GR counterpart exists, and hence it is quite useful to distinguish GR from various versions of STTs. As explained in Section \ref{sec:expmassdist}, the method to measure self-gravitating effects is in generalizing the coupling coefficients $\alpha_a(\varphi)=\partial \log(A(\varphi))/\partial \varphi^a$ given in Eq. \eqref{eq:alpha_a_eq}, to
            \begin{equation}\label{eq:alpha_a_A}
                \alpha_a^A(\varphi) = \frac{\partial \log (m_A(\varphi))}{\partial \varphi^a} = \alpha_a(\varphi) + \frac{\partial \log (\widetilde{m}_A(\varphi))}{\partial \varphi^a} \,,
            \end{equation}
            for the physical Jordan frame masses $\widetilde{m}_A(\varphi) = A^{-1}(\varphi) \, m_A(\varphi)$ and each self-gravitating compact object $A$. Hence, the magnitude of the coefficients $\alpha_a^A(\varphi)$ actually captures the coupling strength of the self-gravitating forces of a compact object $A$ to the multiple scalar fields. This is manifested in the wave equation \eqref{eq:wave_eq_phi}, as to lowest order it yields
            \begin{equation}
                \Box \varphi^a = - 4 \pi G_\star \sum_{A} \alpha^a_A T_A + \mathcal{O}(\varepsilon) \,,
            \end{equation}
            for the localized version of $T$ at body $A$.
            
            We would like to point out that we follow the approach of \cite{Damour:1992we,Damour:1995kt} and work with $\alpha_a^A(\varphi)$ that is different from the standard sensitivities $s_A$ defined in the Jordan frame formulation of the single scalar field PN approach (see e.g. \cite{Mirshekari:2013vb}) and is actually connected to the scalar charge of the body. The exact relation between $\alpha_a^A(\varphi)$ and $s_A$ is extensively discussed in \cite{Mirshekari:2013vb}. Here we will point out only that $\alpha_a^A(\varphi)$ is proportional to $1-2s_A$ in the case of a single scalar field which means that the standard value of $s_A=1/2$ for a GR black hole translates to $\alpha_a^A(\varphi)=0$.
            
            Let us now discuss these self-gravitating effects in the context of our post-Newtonian analysis here. The explicit contributions of the sensitivities $\alpha_a^A$ to the expanded equation of motion potentials \eqref{eq:aj_0}--\eqref{eq:aj_2.5} is due to the direct dependency of the Einstein frame masses $m_A(\varphi)$ and its derivative on the right-hand side of the TMST equation of motion \eqref{eq:EoM}. All contributions are products and/or covariant derivatives with respect to the target space connection of the above given $\alpha_a^A(\varphi)$, namely
            \begin{subequations}\label{eq:betas}
                \begin{eqnarray}
                    \beta^A_{ab} &=& D_a \, \alpha^A_b = \partial_a \alpha^A_b - \gamma^c_{ab} \, \alpha^A_c
                    \, , \\
                    \beta^A_{abc} &=& D_a D_b \, \alpha^A_c
                    \, .
                \end{eqnarray}
            \end{subequations}
	        Due to the way we have formulated the expanded equation of motion \eqref{eq:aj_0}--\eqref{eq:aj_2.5}, we see that all explicit scalar field contributions get contracted with combinations of the here listed coupling fields. Of course, there are also implicit scalar field contributions in the potentials $N_1, B_1, B_2^{ij}, K_2^i, N_2,$ and $B_2$, but the free target space indices inside those fields get contracted with the Riemannian metric $\gamma_{ab}$ and its Christoffel symbols. Hence, we are able to easily distinguish the scalar field contributions related to self-gravitating effects as the explicit appearances in \eqref{eq:aj_0}--\eqref{eq:aj_2.5}. In the case of non self-gravitating bodies, the physical Jordan frame masses $\widetilde{m}_A(\varphi)$ are then independent of the multiple scalar fields $\varphi^a$ and the body-dependent fields \eqref{eq:alpha_a_A} and \eqref{eq:betas} will reduce to their natural body-independent counterpart. The exact influence of these self-gravitating effects, especially with regards to the dipole-radiation phenomena of TMST, is beyond the scope of this work and will be analyzed much deeper in future work when we tackle gravitational waveforms and scalar flux.
	        
        \subsection{Binary Compact Objects}
            Studying the dynamics of binary compact objects and the observed waveforms its full complexity requires one to derive on the one hand the integrals of motion, as well as the equation of motion in the center-of-mass frame, and on the other to derive the expansion of the fields in the radiation zone that is a work in progress. Here we will discuss, though, some conclusions that can be drawn from the equation of motion presented in this paper.
        
        \subsubsection{Binary Black Holes}
            It is a well-known fact that black holes in single scalar field theories obey no-scalar-hair theorems that cover a wide range of possibilities (see e.g. \cite{Herdeiro:2015waa} and references therein). Their PN dynamics is also indistinguishable from GR at least up to 3 PN order. The nonlinear numerical simulations of binary black hole mergers confirm this also for regimes beyond the validity of the PN approach \cite{Healy:2011ef}. If we consider nonrotating black holes similar conclusions will also be true in TMST \cite{Doneva:2020dji}. Using the results in the present paper we can study whether the dynamics of binary black hole systems will also converge to GR if the conditions of this no-hair theorem in \cite{Doneva:2020dji} are satisfied. If we assume that the scalar field is a constant (or zero) and the  black hole mass is independent of the scalar field, then $\alpha_a^A(\varphi)$ and its derivatives are zero.
            If one examines closely the different terms entering the equation of motion  \eqref{eq:aj_0}-\eqref{eq:aj_2.5}, it is clear that the scalar field contribution will be held in $\sigma_\varphi^a$, similar to the single scalar field case \cite{Mirshekari:2013vb}, that are the sources of the field equations of the multiple scalar fields. The explicit form of these sources written in terms of  $\alpha_a^A(\varphi)$ and its derivatives is given in \eqref{eq:rhoasource}. Clearly, zero $\alpha_a^A(\varphi)$ would lead to vanishing $\sigma_\varphi^a$. Therefore, the motion of bald black holes in TMST will coincide with GR at least up to the considered 2.5 PN order.

            The extension to multiple scalar fields, though, brings not only a higher degree of complexity in the equation of motion, but it offers possibilities for new phenomenology. Namely, it is possible to violate the no-scalar hair theorem and produce rotating black holes in TMST with nonzero scalar field \cite{Collodel:2020gyp} that is not allowed in the single scalar field case. The scalar field should have a nonzero scalar field potential, though, that is beyond the studies in the present paper and is a topic of a future work (we refer the reader to \cite{Alsing:2011er,Sagunski:2017nzb} for calculations in the single scalar field case performed though to a lower PN order).
            
        \subsubsection{Binary Neutron Stars}
            Neutron stars, unlike black holes, can easily develop scalar hair in modified gravity since the matter has a nonzero trace of the energy momentum tensor and thus acts as a scalar field source. Again this is encoded in the PN formalist through the quantities $\alpha_a^A(\varphi)$ and its derivatives. What is interesting in TMST is that we can have a set of $\alpha_a^A(\varphi)$ associated with every scalar field that can be significantly different depending on the target space manifold and its metric. This will clearly lead to very interesting possibilities. For example in TMST there exist topological and scalarized neutron star solutions having nonzero scalar hair with a vanishing scalar charge \cite{Doneva:2019ltb,Doneva:2020afj}. It will be interesting to see how the resulting waveforms differ from GR that can be done once we develop the PN formalist in TMST in the radiation zone.
            
        \subsubsection{Black Hole -- Neutron Star Dynamics}
            First let us limit ourselves to the case of nonrotating black holes with zero  $\alpha_{a\; {\rm BH}}^A(\varphi)$ while we allow for a nonvanishing $\alpha_{a\; {\rm NS}}^A(\varphi)$ for the neutron star. In the single scalar field case it was argued that the equation of motion at least up to 3 PN order depend on only a single combination of parameters involving the $\alpha_{a\; {\rm NS}}^A(\varphi)$ \cite{Mirshekari:2013vb, Bernard:2018hta, Lang:2013fna}. This dependence appears in such a way, that it is effectively impossible to distinguish Brans-Dicke theory from other single scalar field theory on the basis of mixed black hole-neutron star binary observations up to this PN order. Such a statement is not true, though, for the general case of TMST because we have an additional stricture that is the target space equipped with a nontrivial metric $\gamma_{ab}$. As extensively discussed above, this metric and its first and second derivatives enter the PN expansion in a nontrivial way that is one of the main qualitative differences between the TMST and the single scalar field. The detailed analysis of the two body equation of motion and the related conserved quantities will be the topic of the second publication of this series. The basis of the calculations performed in the present paper, though, is that one can conclude that the dynamics of a black hole-neutron star system will depend on the particular TMST under consideration, at least for a proper nontrivial choice of the target space metric. Thus the GW observations of such systems can help us discriminate between different subclasses of TMST.
            
            We should, of course, always keep in mind, that if one allows for rotating black holes in TMST, the scalar field and thus $\alpha_{a\; {\rm BH}}^A(\varphi)$ can be nonzero leading to a much more complicated and rich dynamics compared to the single scalar field case.
            
    \subsection{Outlook}\label{sec:outlook}
        This work can be considered as a starting point of a series of papers with the end goal to bring the PN approach and the waveform modeling in TMST to a level that is ready to be compared to GW observations. The next steps are to derive the integrals of motion as well as the equation of motion in the center-of-mass frame with the final goal to understand better the binary dynamics in TMST. Another important extension that has to be done on the way to a proper waveform modeling is the expansion of the fields in the radiation zone. Last but not least, the contribution of nonzero scalar field potential has to also be taken into account. 
        
  	\section*{Acknowledgements}
        We are grateful to Stoytcho Yazadjiev and Carla Cederbaum for helpful discussion and providing feedback on the manuscript. We acknowledge financial support via an Emmy Noether Research Group funded by the German Research Foundation (DFG) under Grant No. DO 1771/1-1. D.D. is indebted to the Baden-Wuerttemberg Stiftung for the financial support of this research project, a Cooperation grant within the framework of the Eliteprogramme for Postdocs. Networking support by the COST actions CA15117 and CA16104 is gratefully acknowledged.

	%%%%%%%%%%%%%%%%%%%%%%%%%%%%%%%%%%%%%%%%%%%%%%%%%%%%%%%%%%%%%%%%%%%%%%%%%%%%%%%
	
	\bibliographystyle{apsrev4-2}
	\bibliography{references}

\end{document}